\documentclass[12pt,preprint]{aastex}

\usepackage[english]{babel}
\usepackage[utf8x]{inputenc}
\usepackage[T1]{fontenc}
\usepackage{amsmath}
\usepackage{amssymb}
\usepackage{graphicx}
\usepackage[colorlinks=true, allcolors=blue]{hyperref}
\usepackage{booktabs}
\usepackage{ulem}

\shorttitle{Habitable Trojans of {\it Kepler} Circumbinary Planets}
\shortauthors{Sudol \& Haghighipour}

\begin{document}

\title{On the Detection of Habitable Trojan Planets in the {\it Kepler} Circumbinary Systems}

\author{Jeffrey J. Sudol\altaffilmark{1} and Nader Haghighipour\altaffilmark{2}}

\altaffiltext{1}{Department of Physics \& Engineering, West Chester University, 720 S. Church Street, West Chester, PA 19383, USA}
\email{jsudol@wcupa.edu, nader@psi.edu}
\altaffiltext{2}{Planetary Science Institute, 1700 East Fort Lowell, Tucson, AZ 85719, USA} 
%\altaffiltext{3}{Institute for Astronomy, University of Hawaii-Manoa, 2680 Woodlawn Dr., Honolulu, HI 96822, USA}

\begin{abstract}
We present the results of a study of the prospect of detecting habitable Trojan planets in the {\it Kepler} 
Habitable Zone circumbinary planetary systems (Kepler-16, -47, -453, -1647, -1661). We integrated the orbits of 10,000 
separate $N$-body systems $(N=4,6)$, each with a one Earth-mass body in a randomly selected orbit near the $L_4$ and
$L_5$ Lagrangian points of the host HZ circumbinary planet. We find that stable Trojan planets are restricted to a 
narrow range of semimajor axes in all five systems and limited to small eccentricities in Kepler-16, -47, and -1661. 
To assess the prospect of the detection of these habitable Trojan planets, we calculated the amplitudes of the variations they
cause in the transit timing of their host bodies. Results show that the mean amplitudes of the transit timing variations 
(TTVs) correlate with the mass of the transiting planet and range from 70 minutes for Kepler-16b to 390 minutes for 
Kepler-47c. Our analysis indicates that the TTVs of the circumbinary planets caused by these Trojan bodies fall 
within the detectable range of timing precision obtained from the {\it Kepler} telescope's long-cadence data.
The latter points to {\it Kepler} data as a viable source to search for habitable Trojan planets.

\end{abstract}

\keywords{Trojan planets (1716), Transit timing variation method (1710), N-body simulations (1083)}

\section{Introduction}

The discovery of circumbinary planets (CBPs) with the {\it Kepler} space telescope has opened new directions 
in the search for habitable planets. Five of the {\it Kepler} CBPs, namely, Kepler-16b 
\citep{Doyle11}, Kepler-47c (the outermost detected planet in a three-planet system) \citep{Orosz19},  
Kepler-453b \citep{Welsh15}, Kepler-1647b \citep{Kostov16} 
and Kepler-1661b \citep{Socia20} reside in the Habitable Zones (HZ) of their host binaries. All these planets 
are larger than Neptune and, therefore, cannot be habitable. However, the possibility remains that these systems 
might harbor habitable moons \citep{Quarles12} or Trojan/co-orbital planets. 

The concept of Trojan and co-orbital planets has long been a subject of interest. 
Prior to the detection of extrasolar planets, research on this topic was limited only to hypothetical systems.
However, the discovery of a large number of planetary systems with a wide array of architectures has changed the 
direction of the research to consider systems beyond what we previously imagined. For instance, 
the detection of two giant planets in
a 2:1 mean-motion resonance (MMR) around the star GJ 876 \citep{Marcy01} demonstrated that
planets may exist in orbital resonances, suggesting that 1:1 MMRs may also be possible.
The latter triggered a wave of studies on the formation, dynamical evolution, and the 
possibility of the detection of Trojan and co-orbital planets. 

The first of such studies was carried out by \citet{Laughlin02}. These authors described three different 
orbital configurations of equal-mass planets in 1:1 MMR that are stable on timescales comparable 
to stellar lifetimes.  \citet{Cresswell06}, using a combination
of hydrodynamical and $N$-body simulations, showed that migrating and still-forming
planets may be captured in 1:1 MMRs and settle into stable tadpole orbits. \citet{Beauge07} carried out
$N$-body simulations of the last stage of the formation of terrestrial planets in the Lagrange points of a 
giant planet and found that the process of the accretion and growth of planetesimals to Earth-sized objects 
does not proceed efficiently. These authors showed that only in rare instances, bodies of
50\% to 60\% the size of Earth may form in the Lagrange points of giant planets. \citet{Lyra09} extended 
these simulations to smaller sizes and showed that, when considering a self-gravitating disk of gas and small 
solid particles, planets with masses 
up to 2.6 Earth-masses can form in the Lagrange points of a giant planet through rapid collapse of accumulated 
solid objects.

Subsequent studies focused primarily on the dynamical evolution and stability of co-orbital
planets. For instance, \citet{Hadjidemetriou11} showed that stable orbits in a 1:1 MMR 
form two distinct families of periodic orbits, a planetary one in which two planets orbit their 
central star in co-orbital motions, and a satellite one in which the two planets form a binary 
revolving around the central star. \citet{Giuppone10} studied the stability of two planets 
in a co-orbital configuration and found that, depending on the mass-ratio of the planets and their orbital
eccentricities, asymmetric periodic orbits appear at anti-Lagrange locations. The appearance of
new co-orbital configurations, especially for eccentric planets, has also been demonstrated by
\citet{Leleu18}. 

The stability of Trojan and co-orbital planets has also been studied in the context of binary star systems.
\citet{Quarles12} showed that the Kepler-16 
system can harbor Trojan objects, and \citet{Schwarz15} demonstrated that the two S-type binaries 
HD 41004 and HD 196885 (where a planet orbits one of the stars of the binary) can harbor Trojan objects as well.
\citet{Demidova18a} and \citet{Demidova18b} have noted that low eccentricities for both the binary 
and its CBP favor the formation of co-orbital structures. The models by these authors also show that the formation of a stable 
co-orbital ring is possible for the Kepler-16 system. Similarly, \citet{Penzlin19} found that co-orbital 
planets can have stable configurations in the Kepler-47 system.
Most recently, \citet{Leleu19} studied the dynamics of a pair of migrating
co-orbital planets and showed that depending on the mass-ratio and eccentricities of the planets,
and also depending on the type and strength of dissipative forces, the two planets may reside
at stable Lagrangian points, may undergo horseshoe orbits, or may scatter out of the system.
 
The discovery of circumbinary planets in the HZ of their host binaries, combined with the fact that
planet formation and migration in circumbinary disks follow similar processes as those around single
stars suggests that the results of the above-mentioned studies on the formation and evolution of Trojan planets can be 
extended to circumbinary systems as well. This motivated us to study the possibility of the existence 
of habitable Trojan planets in these systems and examine their detectability. Given the orbital proximity 
of a Trojan planet to its host CBP, and that transit photometry has been the most successful 
technique in detecting planets around binary stars, we use the variations that a Trojan planet may induce 
in the transit timing of its host CBP as a measure of its detectability. 

Multiple authors have shown that measurements of transit timing variations (TTVs) is a viable technique for detecting 
planets \citep{Miralda-Escude02,Agol05,Ford06,Ford07,Giuppone12,Haghighipour13}. 
As shown by \citet{Agol05} and \citet{Steffen05,Steffen07}, amplitudes of TTVs depend on primarily three factors:
the ratio of the mass of the transiting planet to that of the perturber, 
the orbital period of the transiting planet, and the eccentricity of the perturbing body.
TTVs are greatly enhanced when the two planets are near a mean motion resonance 
\citep[see, for instance][]{Haghighipour11} making this technique especially efficient in detecting planets that are 
near MMRs with the transiting body \citep{Agol05,Agol07,Holman05,Steffen05,Heyl07,Haghighipour11,Veras11}. 

Unlike radial velocity observations, TTVs cannot be modeled by a simple linear superposition of solutions. 
Detailed $N$-body integrations are needed to determine the TTV signature in a system using an extensive library of 
hypothetical objects \citep[see, for instance, the catalog of TTVs in the appendix of][]{Haghighipour11}. 
The method is not without its difficulties. For example, \citet{Ford07} noted that large numbers of transits 
are required to resolve the mass-libration amplitude degeneracy when fitting the TTVs.  
\citet{Nesvorny08}, \citet{Nesvorny09}, and \citet{Nesvorny10} developed a fast inversion 
method to find orbital parameters of a perturbing body from TTVs, however, their approach is unable to
remove degeneracies near MMRs. In a limited study of the TTVs of systems with an Earth-mass planet exterior 
to a hot-Jupiter, \citet{Veras11} showed that a large number of transits $(>50)$ are required to obtain 
meaningful limits on the mass and orbital parameters of the perturbing planet. They also found that the shape 
of the TTV curve may be more useful in breaking degeneracies than their magnitude.  

Utilizing TTVs as a method of detecting terrestrial-class Trojans was first suggested by \citet{Ford07}. 
These authors showed that the amplitude of the TTVs induced by such Trojan planets may fall in the range of 
the sensitivity of ground-based telescopes. \citet{Laughlin02} and \citet{Ford06} had also 
studied the possibility of the detection of Trojan planets. However, their studies were mainly focused 
on the detectability of (giant) planets in 1:1 MMRs using the Radial Velocity technique. As promising as
these studies were, no Trojan planet has yet been discovered from the ground. From space, however,
\citet{Nesvorny12} seem to have detected a planetary candidate, KOI-872c, using the TTV
data from the {\it Kepler} telescope, and \citet{Leleu19} recently announced a co-orbital candidate
using the data from {\it TESS}.

In a study on the detectability of Trojan planets, \citet{Haghighipour13} utilized the enhancement of TTVs
near MMRs, and demonstrated that the amplitudes of variations produced by an Earth-mass 
or a super-Earth Trojan on the transit timing of a giant planet may fall in the range of the
sensitivity of the {\it Kepler} space telescope. These authors considered a system consisting of a star,
a transiting giant planet, and a Trojan body, and integrated the system for different values
of the mass of the central star as well as the masses and semimajor axes of the two planets.
Results of the integrations identified regions of the parameter space for which the orbit of the Trojan
planet would be stable. These authors subsequently calculated the amplitudes of the TTVs induced on the giant planet.

In this paper, we follow the methodology presented in \citet{Haghighipour13}, replacing their
central star with a binary star system. We will carry out thorough dynamical studies of Earth-mass Trojan planets
in the systems of Kepler-16, -47, -453, -1647, and -1661, and determine the range of  
orbital elements that render them stable. We will then calculate the amplitude of the
TTVs induced on the CBPs of each of these systems and discuss the prospect of their detection.  

We continue in \S 2 by describing the set up of our $N$-body integrations.
In \S 3, we present the results of these integrations and in \S 4 we calculate the TTVs due to 
the stable Trojan planets. We conclude this study in \S 5 by discussing the implications of our results 
for the detectability of habitable Trojans in these systems and systems alike.

\section{Numerical Set Up}

We integrated the orbits of a large battery of Earth-mass Trojan planets in the {\it Kepler} HZ 
circumbinary systems. Integrations were carried out using the Bulirsch-Stoer integrator
in a version of the popular $N$-body integrator MERCURY that has been developed specifically for integrating
circumbinary planets \citep{Chambers02}. The physical and orbital parameters of the binaries and their CBPs were obtained 
from their discovery papers and are listed in Tables 1 and 2 (see Introduction for references). 
Prior to integrations, orbital parameters were rotated
from the sky plane to the dynamical plane \citep[see][]{Deitrick15}.  

For each system, we considered 10,000 Earth-mass planets\footnote{We note that, while integrating Trojan planets
in the system of Kepler-1647, three computers in our network failed leaving us 188 integrations short of 10,000. 
Because, compared to other systems, the integrations for Kepler-1647 require an order of magnitude more time to complete, 
we did not attempt to restart these integrations. Because we randomized the list of Trojan 
objects before distributing them across our network of computers, we expect this loss to have no significant 
effect on the results.}. Because, as we explain below, the initial orbital elements of these bodies
were assigned randomly, the total population of them at the $L_4$ (leading) and $L_5$ (trailing)
Lagrangian points were almost equal, $5000 \pm 100$\footnote{The margin of error is the square root of the total
number of Trojan planets (${\sqrt {10000}}=100$).}. Table 3 shows the actual number of planets around each CBP.

We randomly generated the initial semimajor axes of these objects using a Gaussian distribution with a 
range of $\pm 4 \%$
around the semimajor axis of their corresponding CBPs. Their initial eccentricities were also randomly generated
between 0 and $\sim 0.3$ using a one-sided Gaussian distribution with $\sigma = 0.1$. Initial orbital
inclinations were chosen from a range of 
0 to $\sim 30^\circ$ and assigned at random following a one-sided Gaussian distribution with $\sigma = 10^\circ$. 
The initial values of the arguments of periastron, longitudes of ascending node, and mean-anomalies of Trojan 
planets were chosen to follow 
a uniform distribution from $0^\circ$ to $360^\circ$. The combination of these three angles were set to place Trojan 
objects between $30^\circ$ and $90^\circ$ on the either side of the planet. Figure 1 shows the spatial distribution 
of these bodies.

\section{Dynamical Analysis}

For each Trojan planet, we integrated the four-body system of binary-CBP-Trojan for 1 Myr
(in Kepler-47, these integrations were carried out for the six-body system of binary-CBPs-Trojan).
The time step of integrations were set to $\sim$1/30th the orbital period 
of the central binary. We stopped an integration when a Trojan planet collided with its host CBP or
one of the binary stars. We also considered a Trojan to have been ejected from the system when its orbital 
period became equal to or larger than $\sim 30$ times the orbital period of its host CBP. 

Figures 2 and 3 show the results. In these figures, the final dynamical state of each Trojan is shown
in term of its initial semimajor axis $(a)$, eccentricity $(e)$, and orbital inclination $(i)$. Points
colored black represent Trojans that collided with their host CBP, blue corresponds to ejection from
the system, and red indicates Trojans that remained in orbit for the duration of the integration.
Table 4 shows the number of Trojan planets in each category.

As expected, collisions between a Trojan planet and one of the stars of the binary are rare (Table 4). That is due to 
the fact that when a planet enters the region of instability around the central binary, the perturbation 
of the two stars will increase its orbital eccentricity causing it to be ejected from the system instead of 
colliding with the binary stars.

Also, as expected, ejection is not a common outcome. That is because Trojan planets were initially distributed 
around the Lagrangian points of each CBP (that is, inside the CBP's potential well) where they are strongly coupled
to their host planet.  The large number of ejections in the Kepler-16 and Kepler-1661 systems is, however, due to the close 
proximity of their CBPs to the inner unstable region of the host binaries. In these systems, slight perturbations 
from the binary increases the orbital eccentricities of Trojan planets (especially those with initially higher 
eccentricities) causing them to gradually enter the binary's instability region where the gravitational effect of the 
binary scatters them to large distances. This can be seen in figures 2 and 3 where the stable region in Kepler-16 and 
Kepler-1661 systems are slightly off-centered toward larger semimajor axes.

As Table 4 shows, collision with the host CBP is the most common outcome. The reason is that
the perturbation of the binary, and in the case of Kepler-47, that of the two inner planets,
increases the orbital eccentricity of the Trojan body, gradually shifting its orbit into a collision course with its closest 
object, its host CBP (recall that integrations were carried out for individual four-body or six-body systems).
It is, therefore, not surprising that Kepler-47c has the highest number of collisions, and Kepler-1661b is the second,
as it has the smallest periastron distance compared to other HZ CBPs.
The number of collisions drops considerably in the Kepler-1647 system where the CBP 
has the largest semimajor axis among all {\it Kepler} circumbinary systems.

One interesting result of our integrations is the final number of stable Trojan planets in each system. As shown in Table 4,
except for Kepler-16b, the number of stable Trojans at the $L_4$ (leading) and $L_5$ (trailing) Lagrangian points of all
other CBPs are statistically identical. However, in the Kepler-16 system, more Trojan planets are stable at $L_5$ than 
around $L_4$. At the first glance, this seems to be 
in contrast to the number of Trojan asteroids at the Lagrangian points of Jupiter where more than 2/3
of these bodies are around the $L_4$. However, it is important to note that a comparison between the Trojan planets
considered here and Jupiter's Trojan asteroids is not entirely valid. The planets considered here strongly affect
the dynamics of their host CBP and must be considered within their full $N$-body system  whereas Trojan asteroids are 
effectively massless particles with no effects on the orbit of Jupiter. The dynamics of these bodies, in addition to 
their interaction with Jupiter, is also heavily affected by non-gravitational forces as well as their mutual collisions,
two factors that do not exist in the our systems. We are currently investigating this asymmetry in Kepler-16 stable Trojans 
and will publish our results in a future article.

Another interesting result, as depicted by figures 2 and 3, is the extent of the region of stability.
In systems where the CBP is close to the binary (Kepler-16, Kepler-1661) and where there are 
other strong perturbations (Kepler-47), 
stable Trojans are limited to a narrow region in semimajor axis and include only those with low eccentricities. However, 
in a system like Kepler-1647 where the CBP is farther out, the stability region expands both in semimajor 
axis and eccentricity. 

Figure 3 also shows a gap in the stable region around Kepler-16b where, for a small range of inclination, the orbit of the 
Trojan planet becomes unstable. We speculate this island of instability to correspond to unstable inclination resonances 
between the Trojan planet and the central binary. Investigation on this topic is currently underway and will be published elsewhere.

The extent of the stability region with regard to orbital eccentricity raises the question whether 
Trojans with higher eccentricities will be stable if integrations are continued for longer times
(e.g., 10-100 Myr). While to answer this question, 
continuing integrations seems to be the natural choice, in practice, the short periods of the
binaries considered here (periods vary from 7 days for Kepler-47 to 41 days for Kepler-16) 
combined with small time steps (1/30th of the period of the binary) make
the extension of integrations to longer times prohibitive. For this reason, 
instead of direct integration, we used the value of the maximum eccentricity $(e_{\rm max})$ of each Trojan as a measure of
its long-term stability. In general, the value of $e_{\rm max}$ allows the identification and exclusion of regions
of the parameter space where the orbit of an object becomes parabolic (unstable). The lower the value of $e_{\rm max}$, 
the longer the Trojan will maintain its orbit. Figures 4 and 5 show $e_{\rm max}$ for all Trojans in our systems in terms of
their initial eccentricities and orbital inclinations. 
As shown in these figures, the range of the semimajor axes of Trojans with the lowest $e_{\rm max}$ varies
between $\pm 1\%$ and $\pm 3.5\%$ of the range of the semimajor axes of their corresponding CBPs. In the next section,
we consider Trojan planets within this range for calculating TTVs.

We would like to note that, because in different systems, transits occur at different times, a more appropriate
way to compare the dynamical states of the systems would be to integrate them for the same number of
transits rather than a fixed amount of time (e.g., 1 Myr). For instance, given that in the systems 
studied here, the orbital period of Kepler-1647b is almost three times the orbital periods of other CBPs, it is 
possible that if the four-body systems of binary-CBP-Trojan in Kepler-1647 were integrated for 3 Myr, 
more Trojans would become unstable and their region of stability would become smaller. 
We would, however, like to emphasize that the purpose of this paper is not to study the dynamical evolution 
of {\it Kepler} CBPs and their hypothetical Trojans. The goal of this paper is to assess the detectability 
of such Trojan bodies via the transit timing variation method. Stability analyses are merely carried out to exclude unstable
systems from the TTV integrations.

\section{Transit Timing Variations}

To calculate the transit timing variations of a CBP due to the effect of a Trojan planet, the orbit of the CBP 
must be integrated while subject to the perturbation of the Trojan. Although, intrinsically, these integrations are 
similar to those carried out for the stability analyses, it is not possible to use the results of the stability integrations 
to calculate the TTVs. The reason is that TTV integrations require much smaller time steps to resolve the orbital 
interpolations necessary for calculating variations in the orbit of a CBP.

Before explaining the TTV integrations, we would like to note that it is unnecessary to carry out these integrations 
for a long time. Because the purpose is to calculate the amplitude of TTVs, an integration time to produce a few TTV 
cycles will suffice.
Also at this stage, it is not necessary to carry out integrations for all stable Trojans. Many of these bodies are close
to one another in the parameter-space and as a result, the amplitudes of the TTVs they produce will be close as well.
A more efficient approach is to integrate only a sample of them.

To construct such a sample, we divided the population of stable Trojans in each system into 18 zones: three in radius 
from the central binary and six in angle from the CBP with three leading and three trailing the planet. We then randomly chose 
300 Trojans from the 18 zones in each of the systems of Kepler-16, Kepler-453, and Kepler-1647, 200 from the system 
of Kepler-1661 and 100 from the Kepler-47 system. The smaller sample size in Kepler-1661 is the result of its smaller population
of Trojans. In the case of Kepler-47, not only is the population smaller, but also the integrations take considerably more time.

We integrated the four-body system of Binary-CBP-Trojan (six-body in the case of Kepler-47) 
for each Trojan planet in the sample and for the duration of two complete TTV cycles 
with a time step of 0.001 days. This integration time translates to 36 transits 
at the low-end for Kepler-1647 and 280 transits at the high-end for Kepler-47.
We assumed that at $t=0$, the primary star was at the origin of coordinate 
system and placed the CBP on  the $x$-axis with the positive direction of the axis toward the observer. We determined 
the times of mid-transits of the CBP by interpolating between the times before and after its center crossed 
the line of sight to the center of the star.

Figures 6-8 show samples of the results. Because the mass of the inner planet in the Kepler-47 system 
is not well constrained, we calculated two sets of TTVs for this system, one with the inner planet being
2.073 Earth-masses and one with its highest possible mass, 25.77 Earth-masses \citep{Orosz19}. Also, because
the variations in the orbit of Kepler-47c is due to the combined effects of the two inner planets
and the Trojan body (it is not possible to separate these effects), we fit a tilted sine function to each TTV, 
and take the amplitude of the fitted function to be the amplitude of the TTV. We note that, as shown 
in figure 8, the effect 
of the mass of the inner planet on the transit timing variations of Kepler-47c is almost negligible. 

Table 5 shows the maximum, minimum and the median amplitudes of the TTVs in all our systems, 
arranged in descending order of CBP mass. A few interesting features appear in this table 
and figures 6-8.

First, as shown by the figures, the amplitudes of TTVs increase with an increase in the range
of eccentricities of the Trojan planet (within the stable region). This is an expected result because
Trojans with larger eccentricities have closer approaches to their host CBPs which enhances the effect of their 
perturbations. 

Second, Table 5 also shows that both the periods and amplitudes of the TTVs 
increase as the CBP mass becomes smaller. This is expected and consistent with the findings of \citet{Agol05},
\citet{Steffen05}, and \citet{Steffen07} who showed that when the ratio of the mass of the perturber to 
the transiting planet increases, the amplitude of TTVs increase. In our study, the perturber is a
planet with a fixed mass of one Earth-mass and, therefore, the ratio of its mass to the mass of the transiting 
planet (i.e., the CBP) increases as the mass of the CBP decreases. 

Finally, a comparison between the amplitudes of TTVs 
in our systems and 
the transit timing variations of a large sample of {\it Kepler} planets \citep[see figure 1 in][]{Ford11} 
indicates that the values of the TTVs of our systems fall squarely in the range of the sensitivity of the
{\it Kepler} space telescope. That means, if Trojan planets can form and remain in stable orbits, {\it Kepler} data 
would provide the best resource to search for their TTV signatures. For instance, the period of secular oscillations  
of the TTVs of Kepler-16b is approximately 12.5 years, which suggests that the orbital characteristics of a 
massive Trojan planet could be inferred from a decade of transit data of this system.

\section{Summary and Concluding Remarks}

We have presented the results of a study of the detectability of Earth-mass Trojan planets as
possible habitable bodies in the {\it Kepler} HZ circumbinary systems. We analysed the stability
of these planets for different values of semimajor axis and eccentricity around the $L_4$ and $L_5$
Lagrangian points of their host CBPs, and identified regions of the parameter space where these planets can
have long-term stable orbits. Results indicated that numerous stable orbits exist for Trojan
planets in these systems. Motivated by the latter, we calculated the variations these Trojans planets create
in the transit timings of their host planets and found that these TTVs fall squarely in the range of the timing precision 
obtained from the {\it Kepler} telescope's long-cadence data. As expected, the prospect of detection is higher in systems 
where the CBP has a smaller mass and for stable Trojan planets with high eccentricities.
As shown by \citet{Haghighipour13}, the results of our study can be used as a pathway to break the strong degeneracy 
associated with the RV-detection of Trojan planets around  $L_4$ and $L_5$ \citep{Giuppone12} by combining the 
information obtained from the study of the transits of their systems and their TTVs.

Although the amplitude of the TTVs are within the range of photometric sensitivity of the {\it Kepler} telescope,
if such TTVs are detected, it is not possible to determine if they are due exclusively to a Trojan planet.
As shown by \citet{Nesvorny08}, \citet{Nesvorny09}, \citet{Nesvorny10}, and \citet{Nesvorny12},
the mass and orbital elements of the perturbing body can be determined by analyzing TTVs alone in non-resonant systems. 
However, to determine these quantities in a near-resonance system, more information is needed. For instance, a catalog
of TTVs, similar to those created by \citet{Haghighipour11} can be used to determine the masses and orbital elements of the 
transiting and perturbing planets, although that approach too suffers from degeneracies. 
Another approach would be a combination of decade-long radial velocity 
and transit timing variation measurements \citep{Laughlin02,Giuppone12}
along with high time resolution, $N$-body integrations to identify non-contributing 
(i.e., unstable) regions of the parameter space.

The premise of our study is the assumption that planetary systems may exist in which planets are in a 1:1 MMR.
Given the diversity of extrasolar planets, and, in particular, the existence of a variety of systems with different 
orbital resonances, it seems plausible that systems with Trojan or co-orbital planets may also exist.
How such planetary systems form,
however, is an important question that requires deep investigation. As mentioned in the introduction, the in-situ formation 
of 1:1 resonant planets is rather unlikely. Planet migration is also destructive and does not allow systems to 
maintain their Trojan or co-orbital orbits. That leaves only one scenario, capture in a 1:1 MMR. Whether capture
occurs during the formation, as a result of planet-planet interaction, and/or because of planet migration is a topic
beyond the scope of this paper and requires more thorough investigations.

As a final note, our $N$-body integrations show no preference for Trojan objects sharing the argument of periapsis 
or the longitude of the ascending node with the transiting planet. Therefore, from a late-stage, dynamical 
perspective and barring any requirements set by the formation mechanisms for co-orbiting planets, 
the probability of detecting a double transit is the square of the probability of detecting the transit of either 
the planet or its Trojan companion. Since the probability of detecting any planet in transit is on the order of 1 
in 100, only 1 in 10,000 systems with a Trojan planet large enough to be detected might be so well aligned as to 
exhibit a double transit. Given that only 13 circumbinary planetary systems have been discovered by the transit method, 
a double transit system would be an extraordinary find.

\acknowledgments
JJS acknowledges support from the West Chester University College of Sciences and Mathematics and the 
NASA Astrobiology Institute. NH acknowledges support from NASA XRP through grant number 80NSSC18K0519. 
We would like to thank the anonymous referee for their critically reading our manuscript and very useful 
recommendations that improved our paper. Most of the $N$-body integrations presented in this study 
were carried out on a large network of computers graciously provided by the Information Services and Technology Department 
at the West Chester University.

\clearpage
\begin{figure}
\vskip -30pt
\centering
\includegraphics[width=0.39\textwidth]{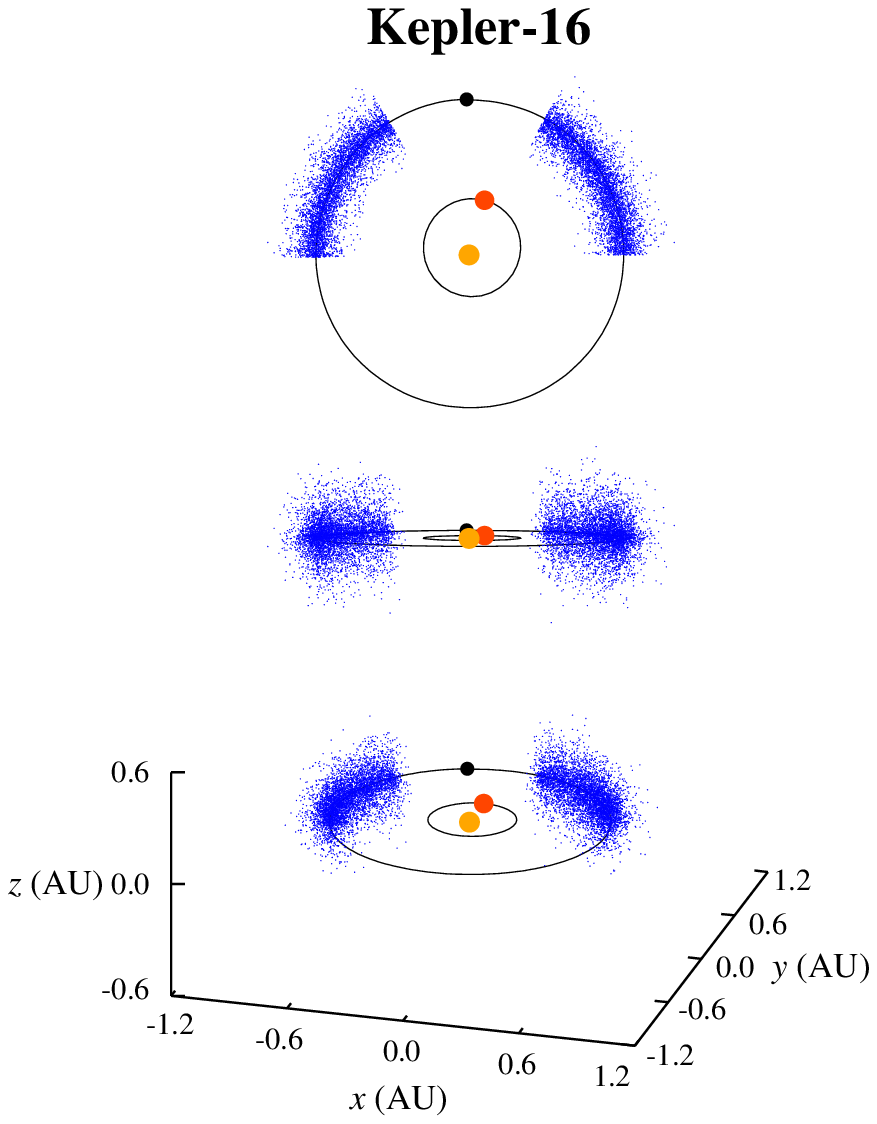}
\hskip 100pt
\includegraphics[width=0.35\textwidth]{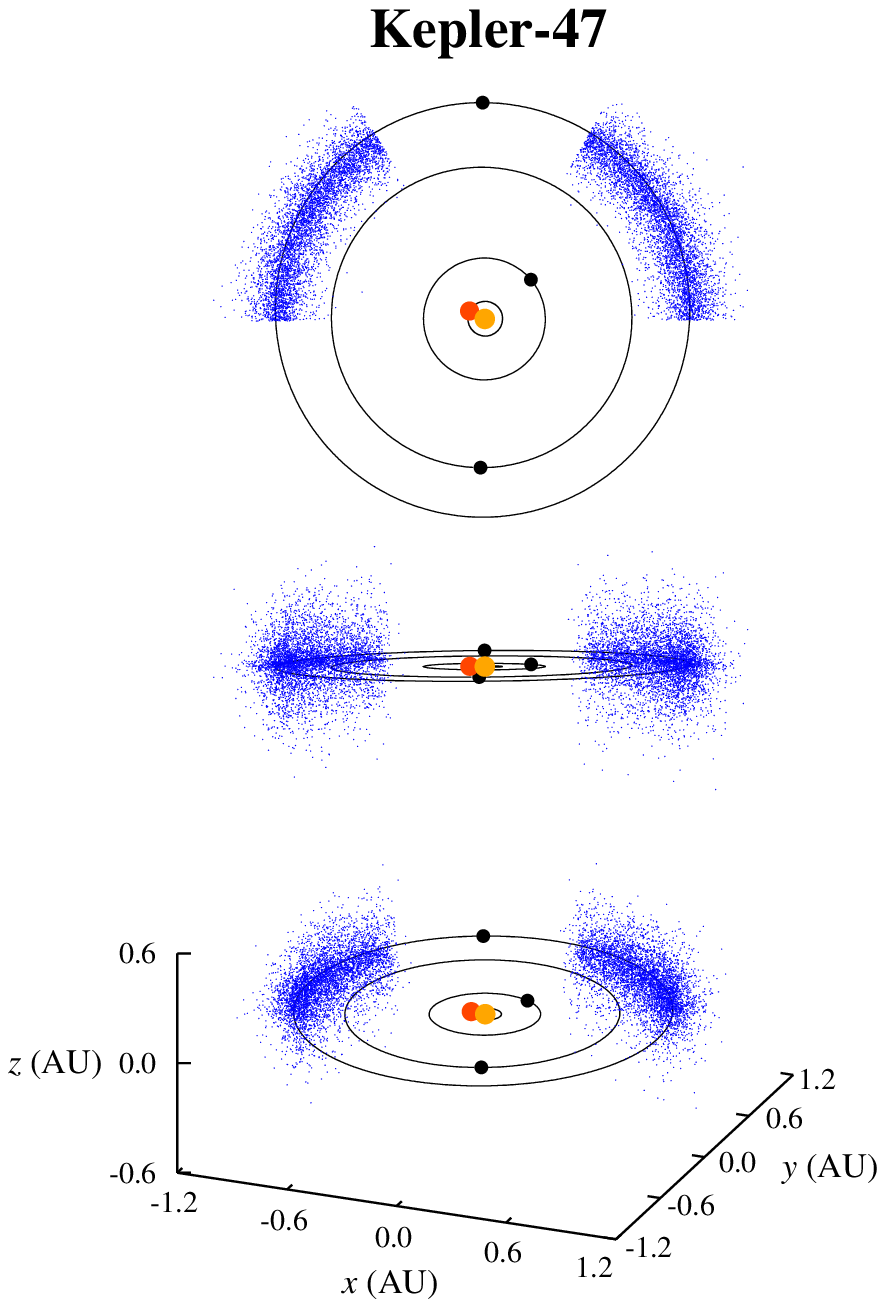}
\center{
\vskip -120pt
\includegraphics[width=0.42\textwidth]{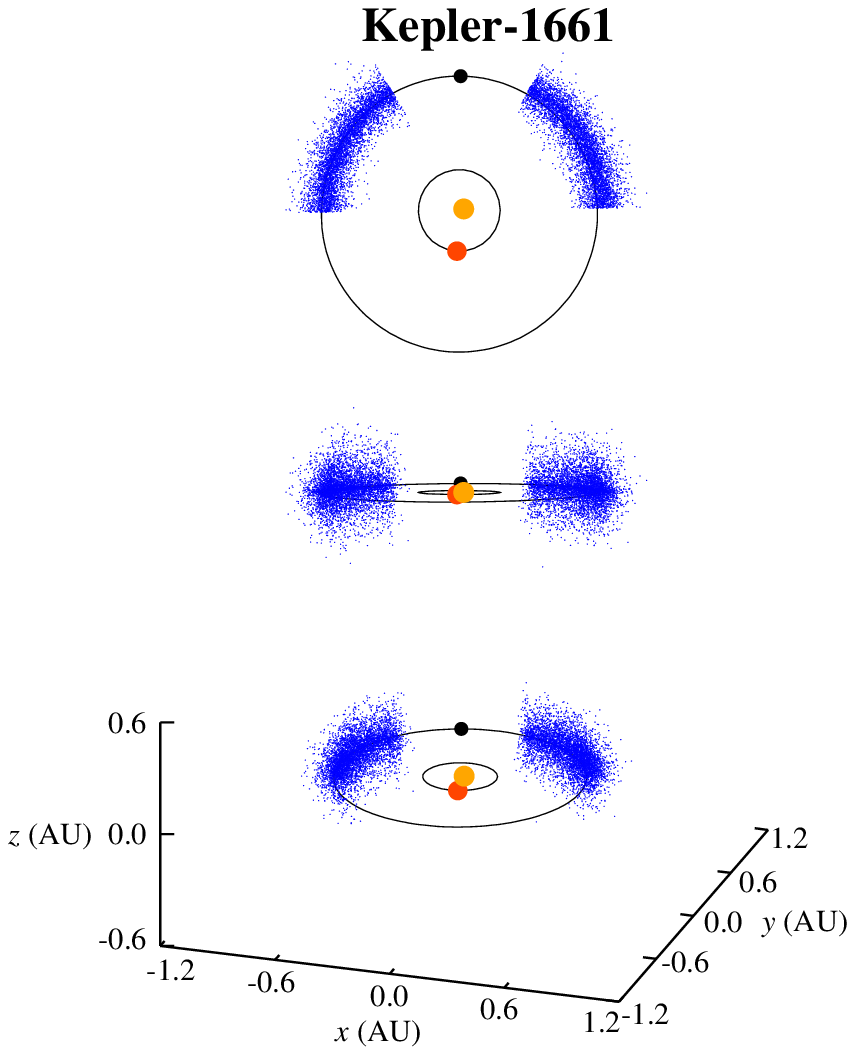}}
\center{
\vskip -80pt
\includegraphics[width=0.41\textwidth]{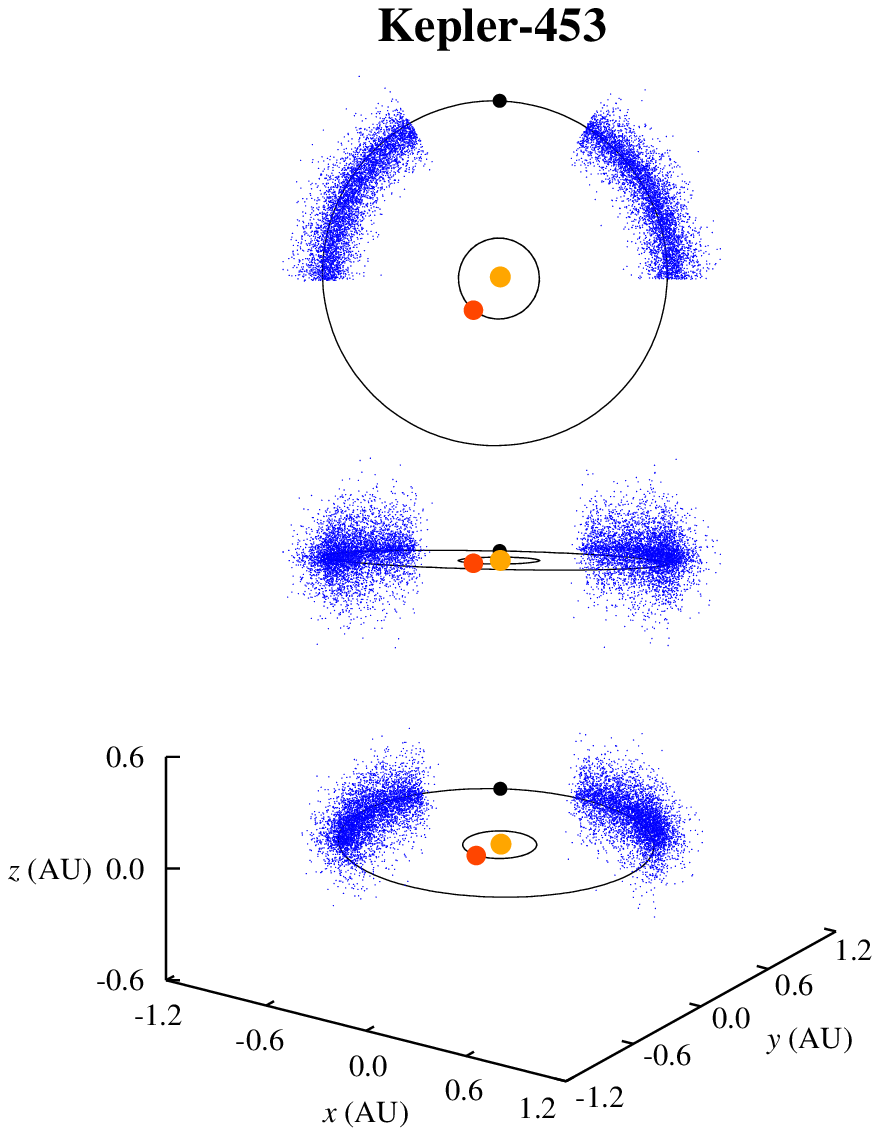}
\hskip 95pt
\includegraphics[width=0.38\textwidth]{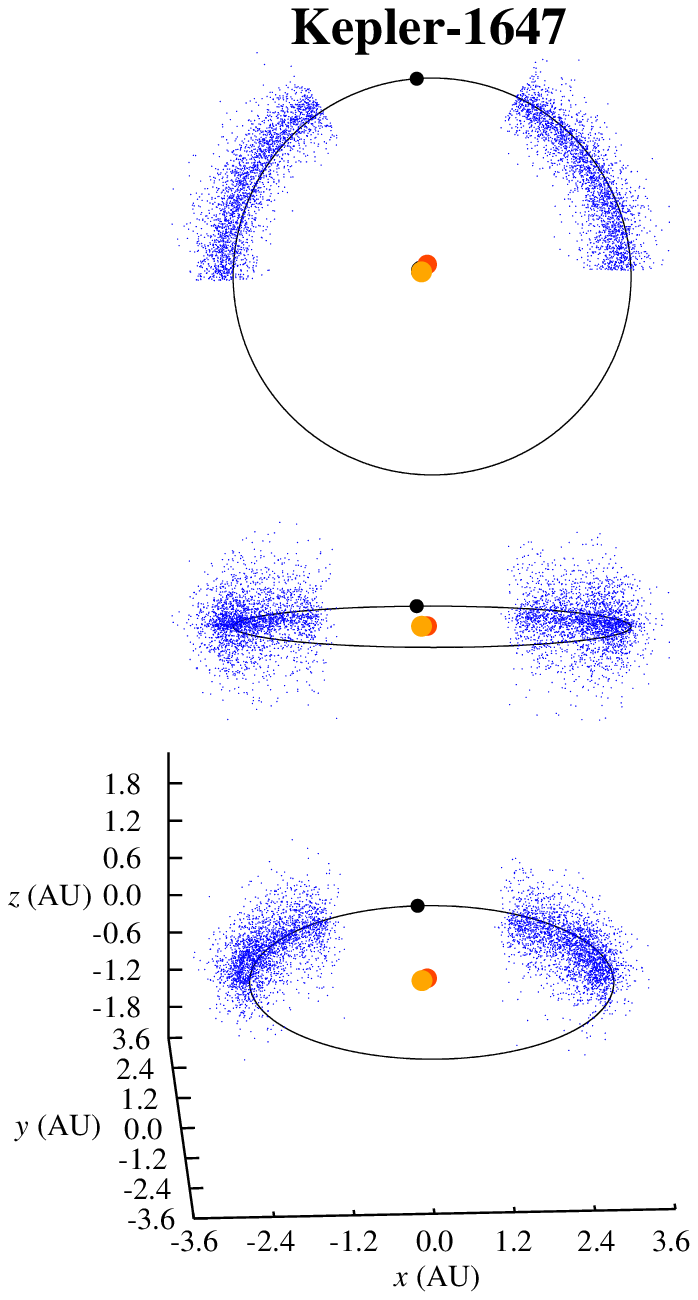}}
\vskip -20pt
\caption{Spatial distribution of Earth-mass Trojan planets in each of the {\it Kepler} 
systems considered here. The orange circle represents the primary star, the
red circle is the secondary, and the black circle is the CBP.
For each system, the top panel shows an overhead view, the middle
panel shows an edge-on view, and the bottom panel shows a view from $68^\circ$ north of equator along line between 
the primary star and the CBP.}

\end{figure}

\clearpage
\begin{figure}[h]
\vskip -10pt
\centering
\includegraphics[width=0.42\textwidth]{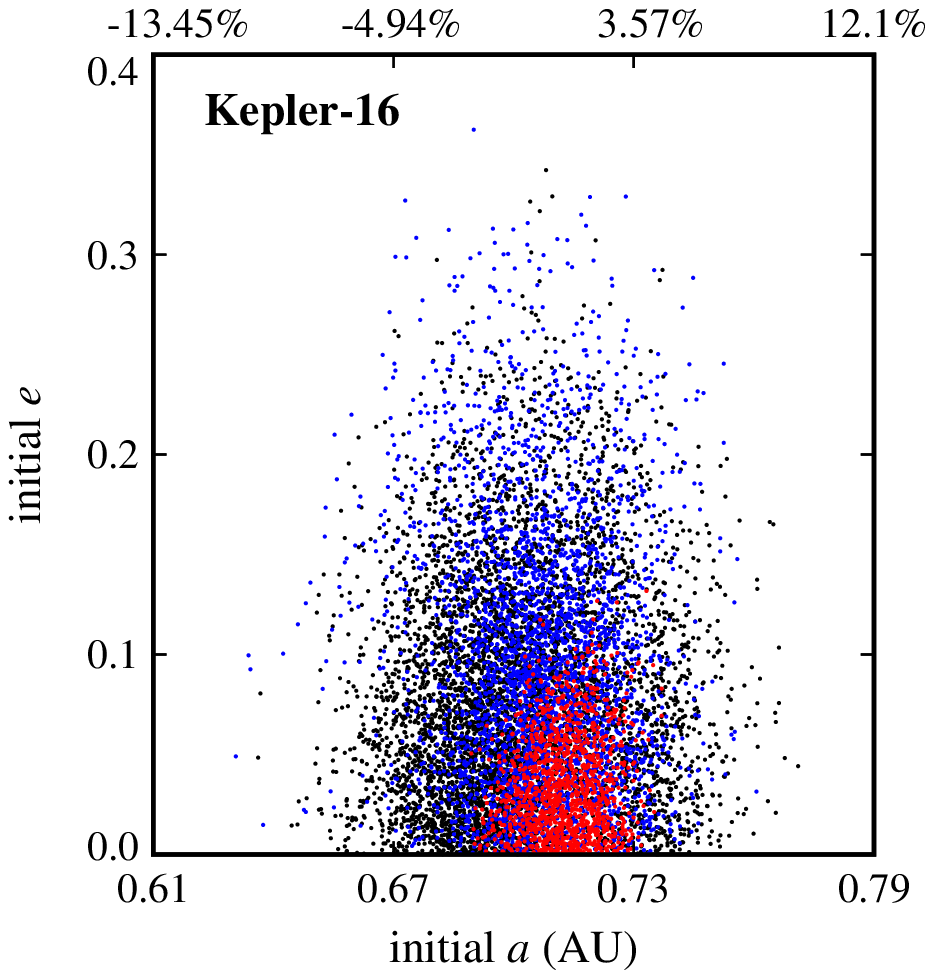}
\hskip 34pt
\includegraphics[width=0.42\textwidth]{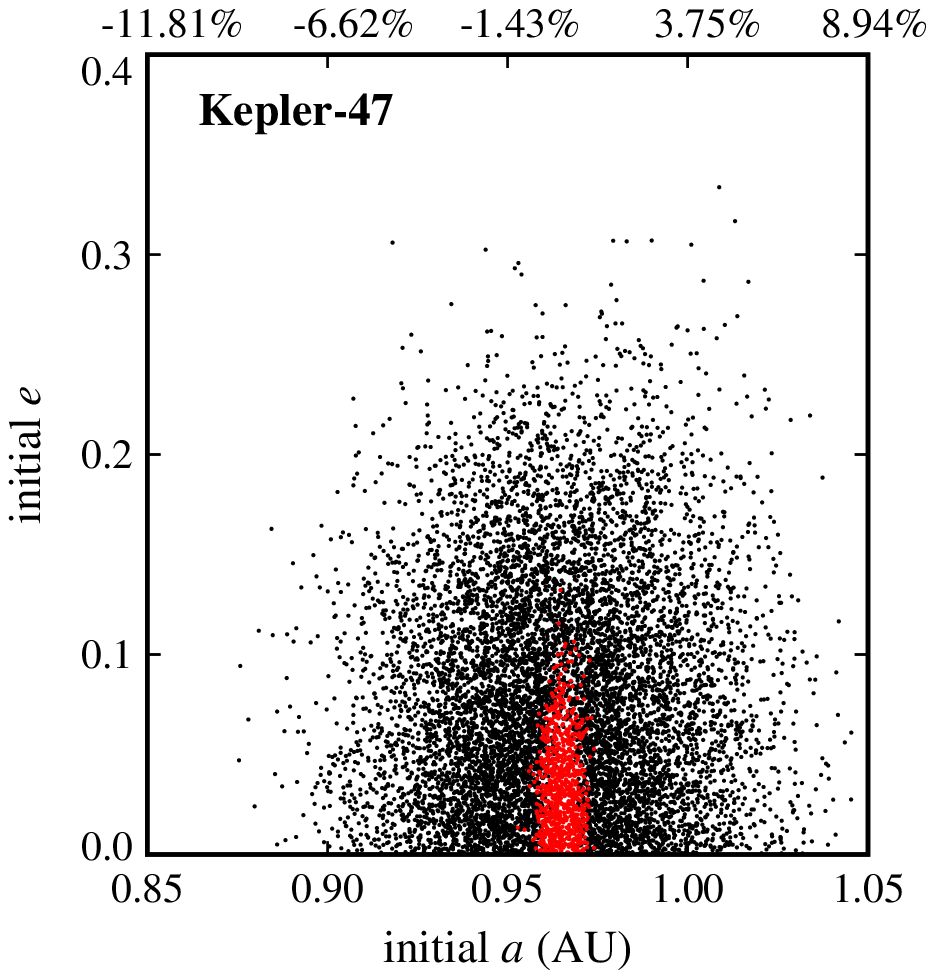}
\center{
\vskip -10pt
\includegraphics[width=0.42\textwidth]{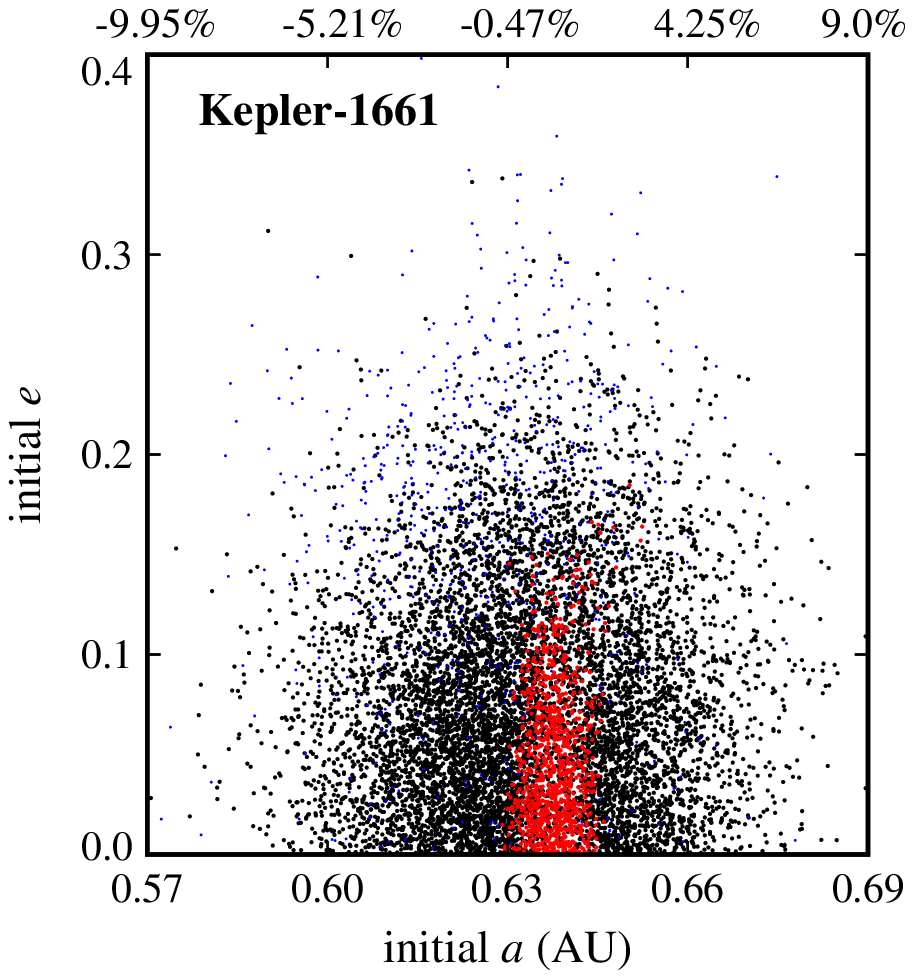}}
\center{
\vskip -10pt
\includegraphics[width=0.42\textwidth]{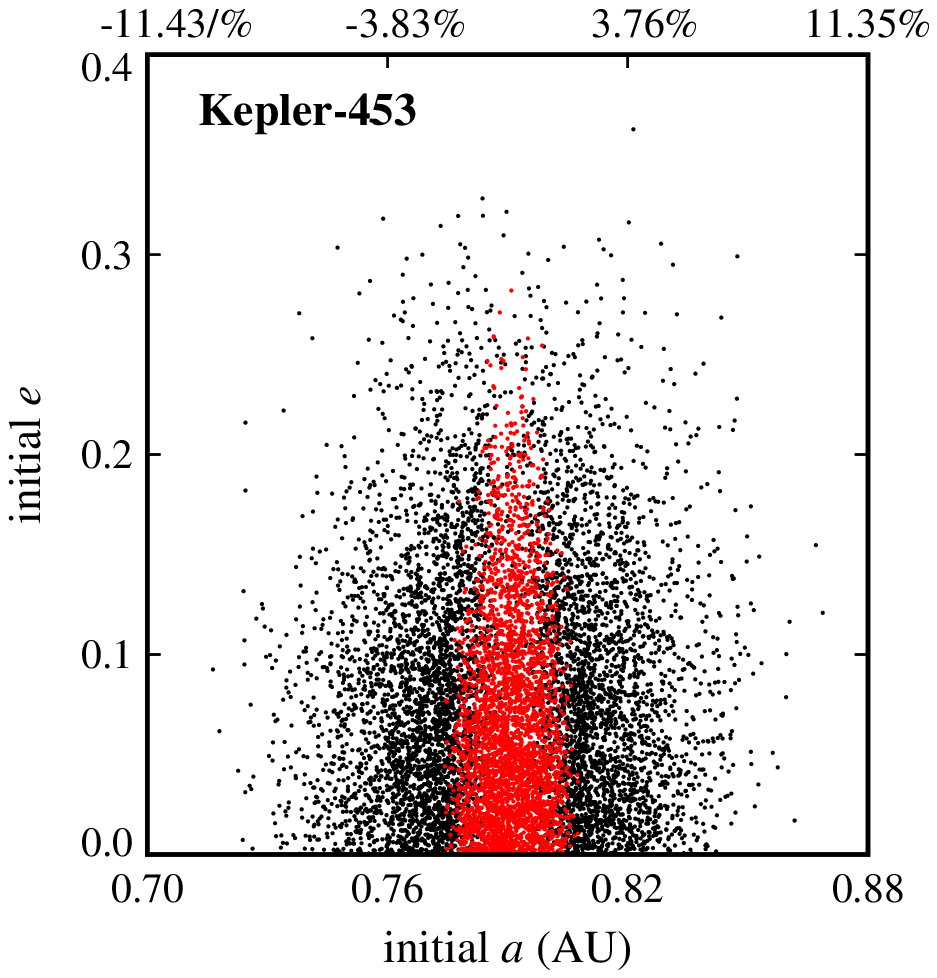}
\hskip 34pt
\includegraphics[width=0.42\textwidth]{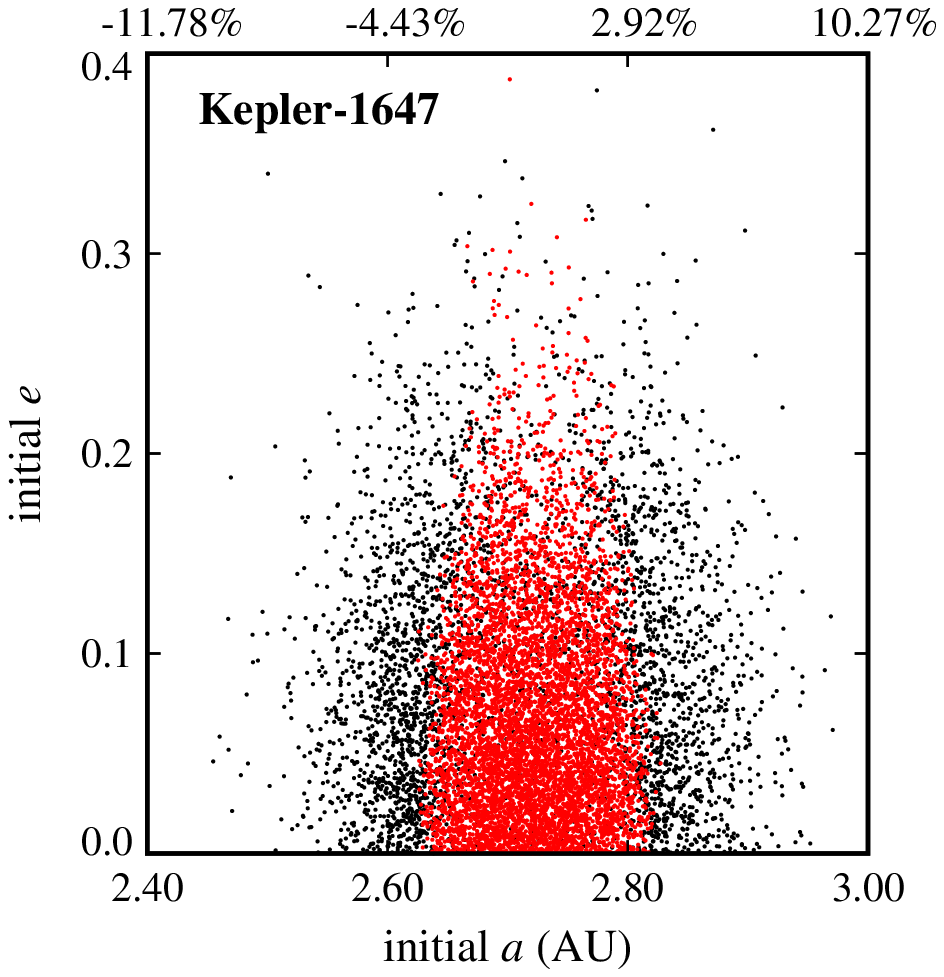}}
\vskip -10pt
\caption{The final dynamical states of the Trojan planets with respect to their initial semimajor axes 
(centred on the CBP semimajor axis) and their initial orbital eccentricities. Black indicates a 
collision with the host CBP, blue indicates ejection from the system, and red indicates the Trojan planet 
remained in orbit until the end of the integration. The top horizontal axis shows the percent deviation 
from the CBP semimajor axis.}
\end{figure}

\clearpage
\begin{figure}[h]
\vskip -10pt
\centering
\includegraphics[width=0.42\textwidth]{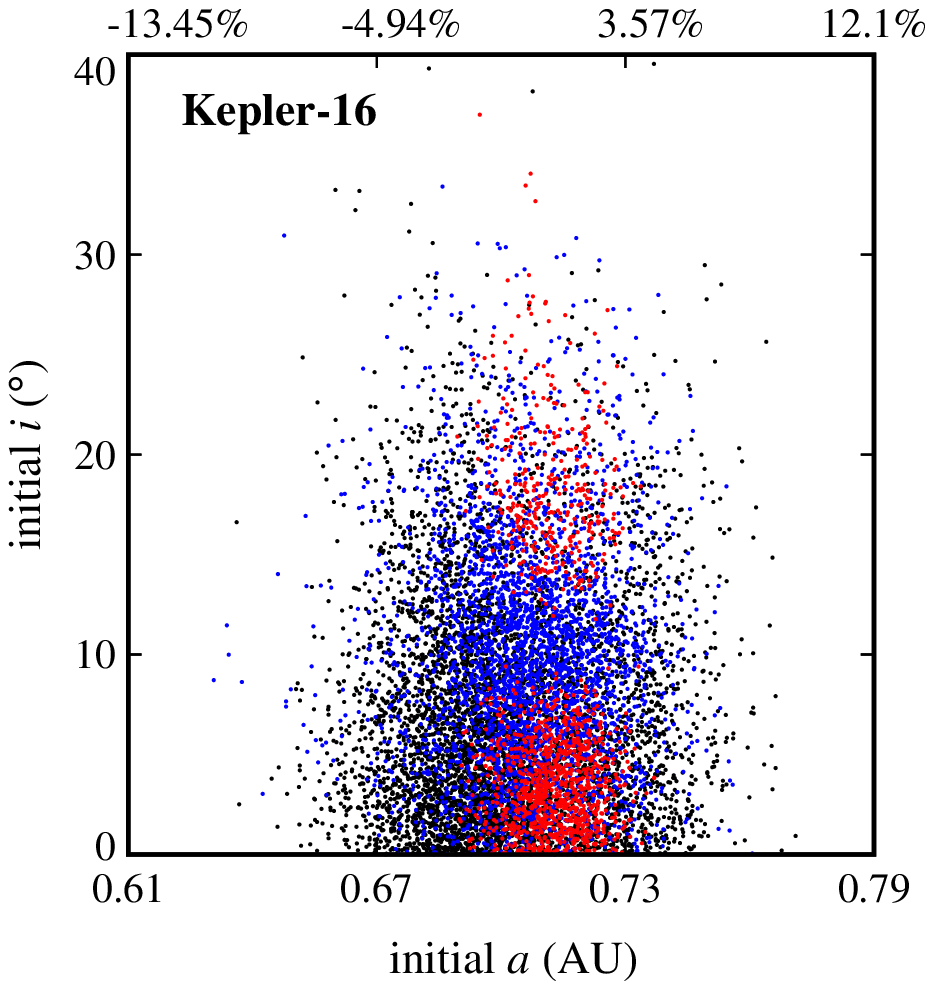}
\hskip 34pt
\includegraphics[width=0.42\textwidth]{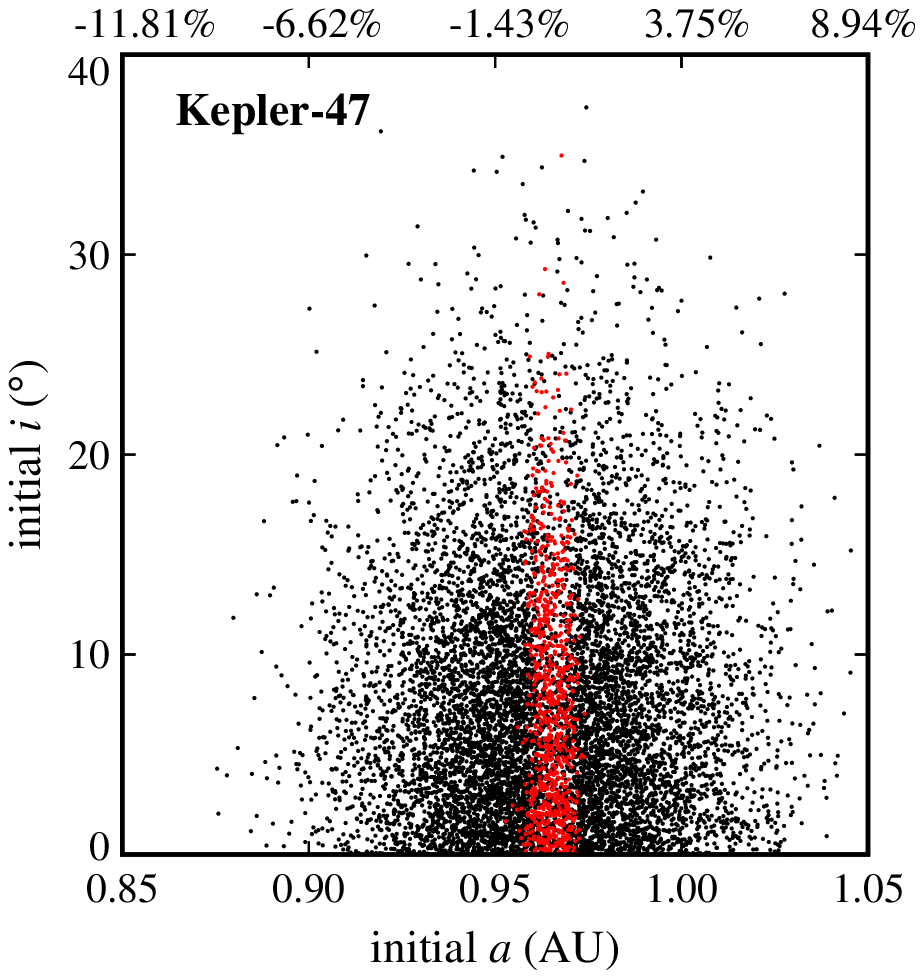}
\center{
\vskip -10pt
\includegraphics[width=0.42\textwidth]{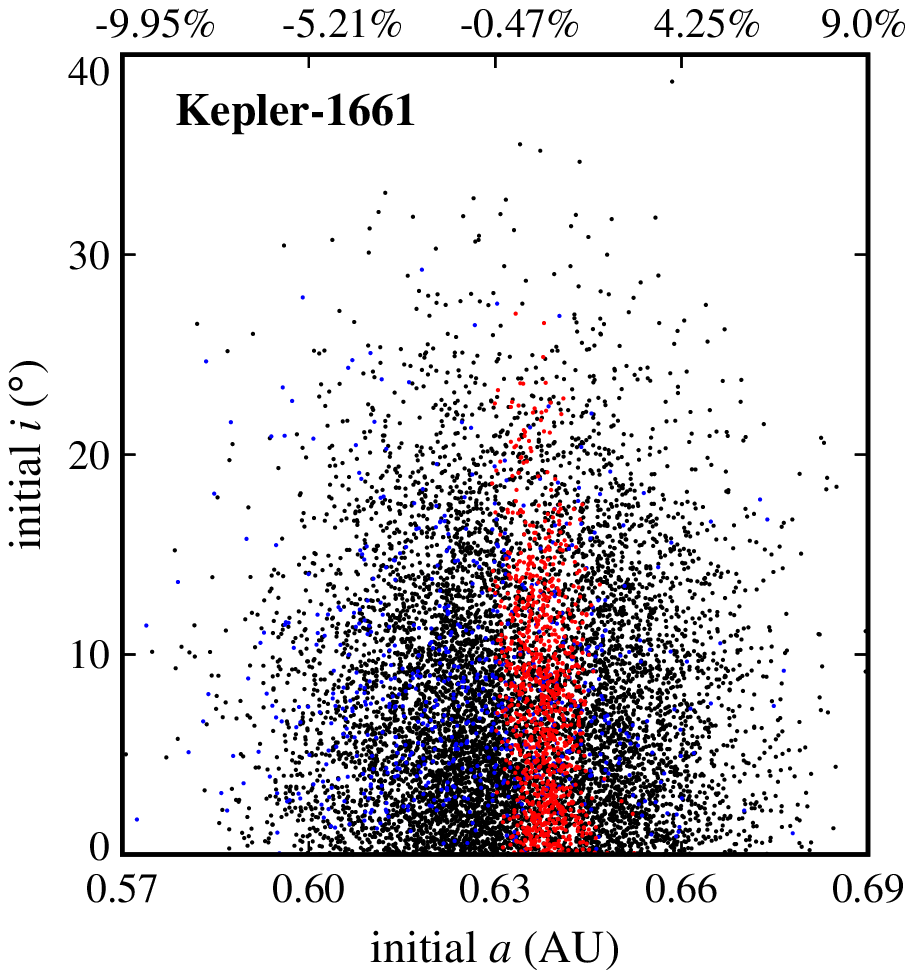}}
\center{
\vskip -10pt
\includegraphics[width=0.42\textwidth]{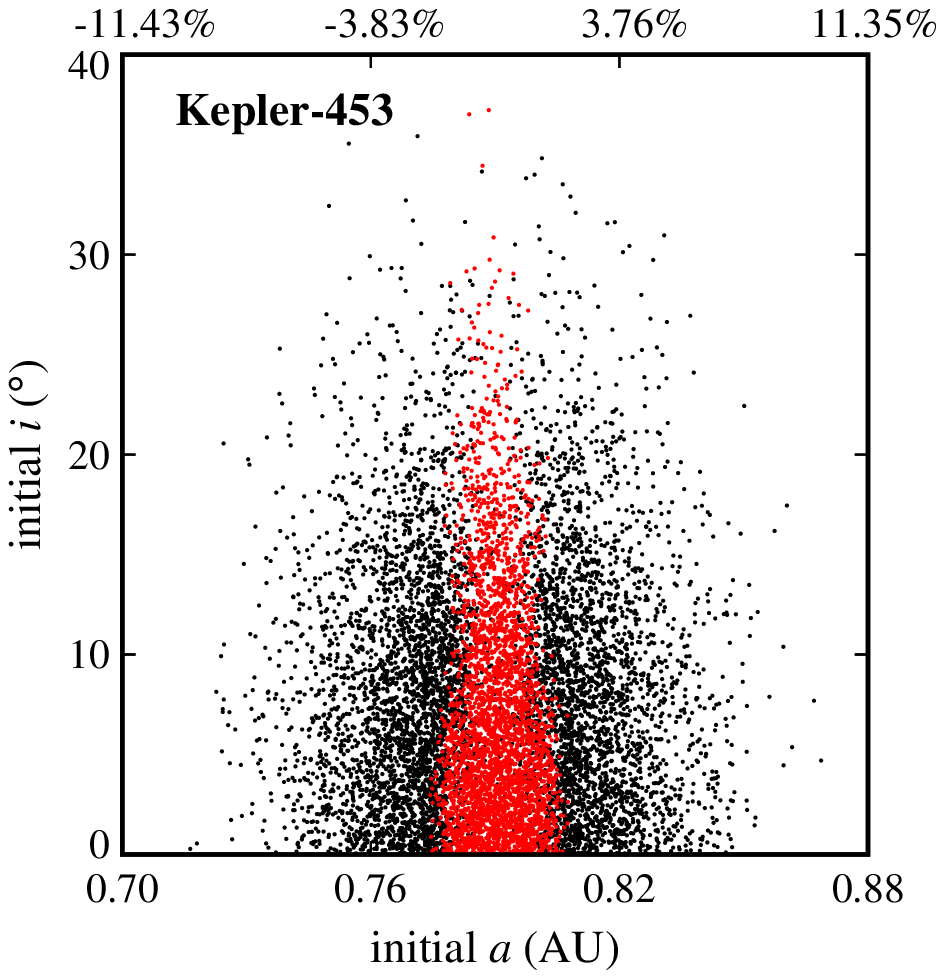}
\hskip 34pt
\includegraphics[width=0.42\textwidth]{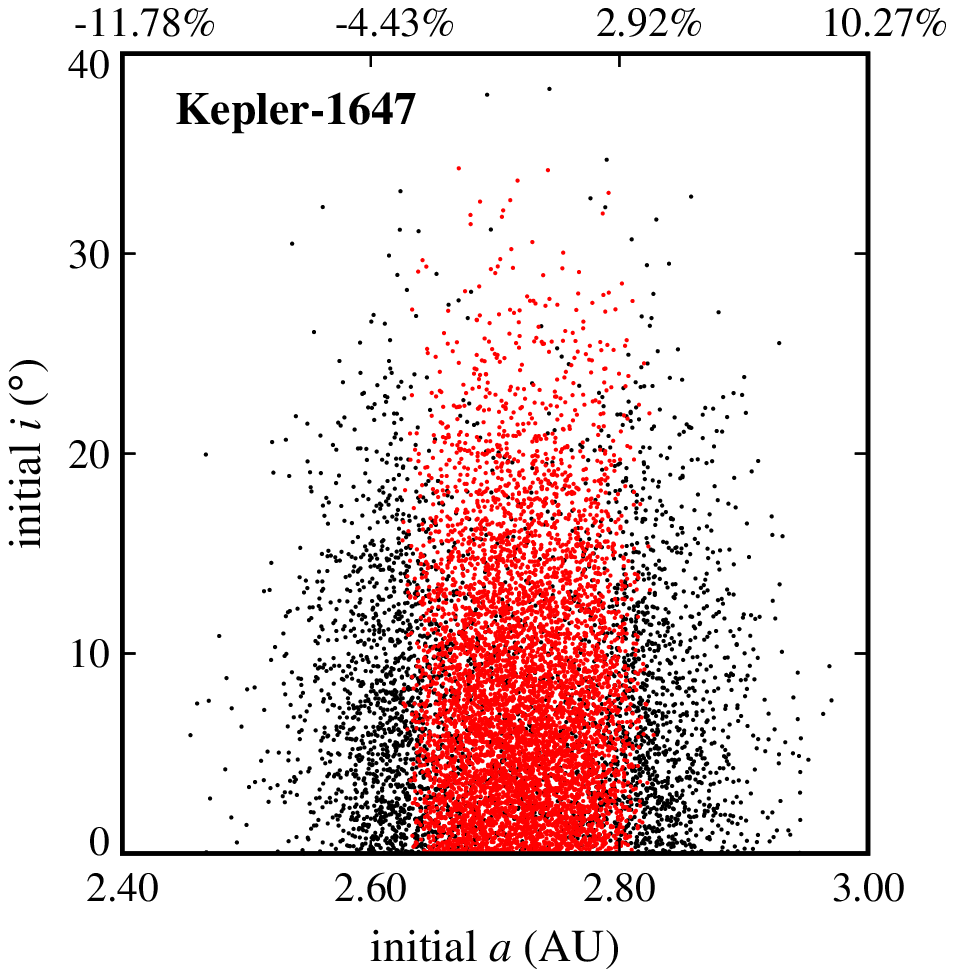}}
\vskip -10pt
\caption{The final dynamical states of the Trojan planets with respect to their initial semimajor 
axes (centred on the CBP semimajor axis) and their initial orbital
inclinations.  Black indicates a collision with the host CBP, blue indicates ejection from the system, 
and red indicates the Trojan planet remained in orbit until the end of the integration.
The top horizontal axis shows the percent deviation from the CBP semimajor axis.}
\end{figure}

\clearpage
\begin{figure}[h]
\centering
\includegraphics[width=0.42\textwidth]{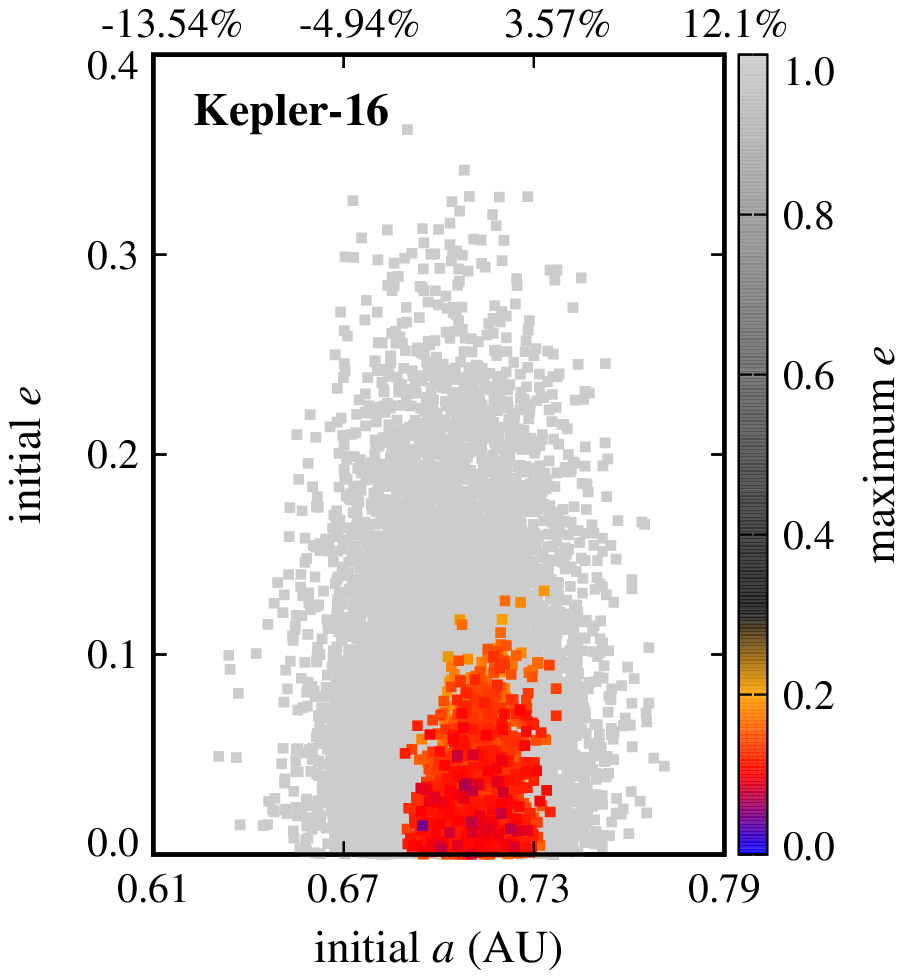}
\hskip 34pt
\includegraphics[width=0.42\textwidth]{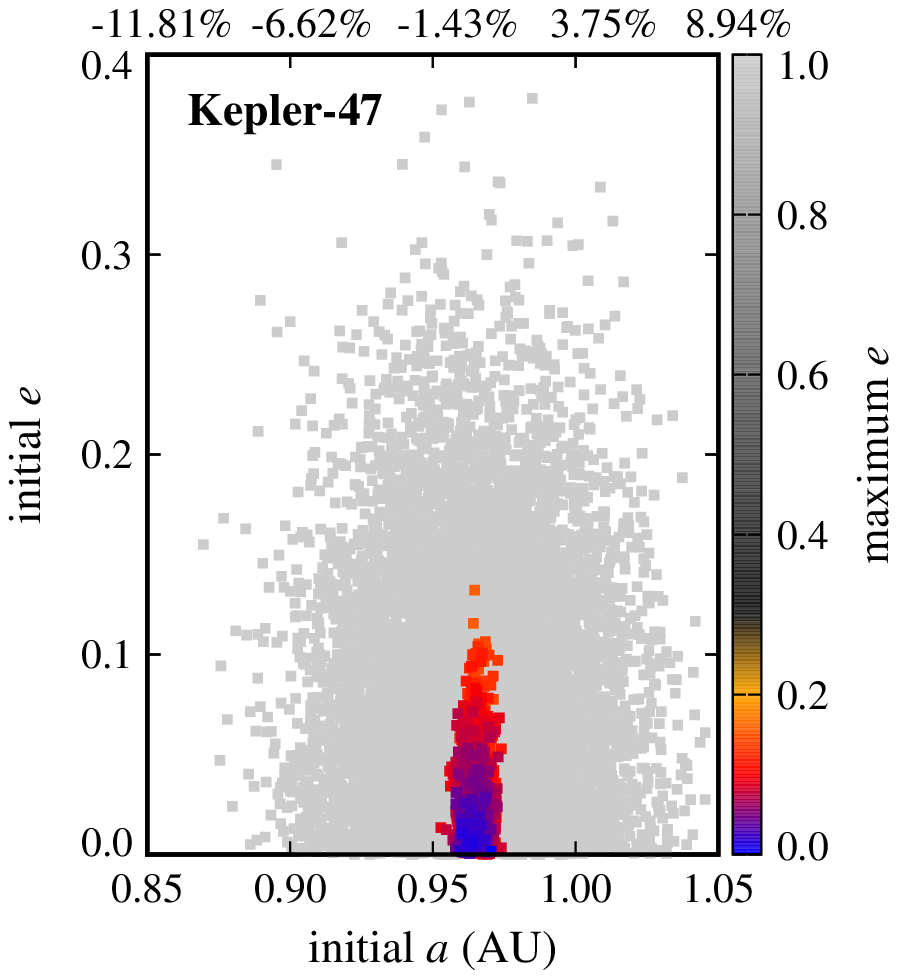}
\center{
\vskip -10pt
\includegraphics[width=0.42\textwidth]{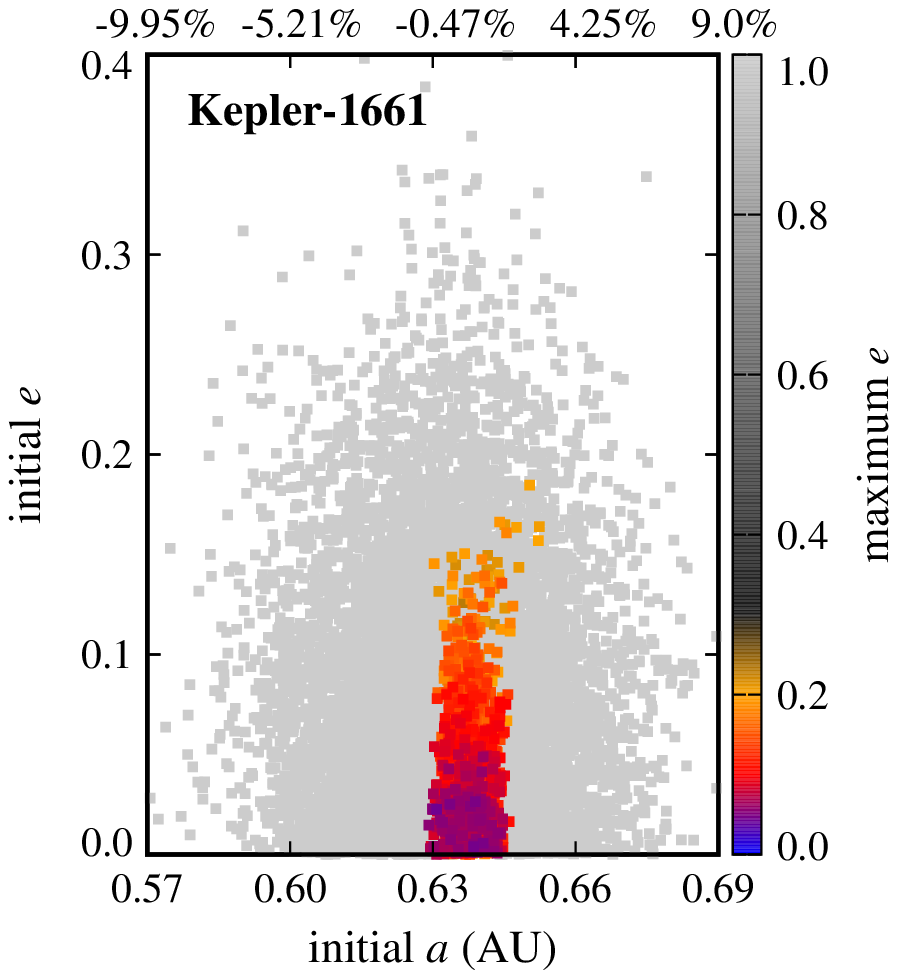}}
\center{
\vskip -10pt
\includegraphics[width=0.42\textwidth]{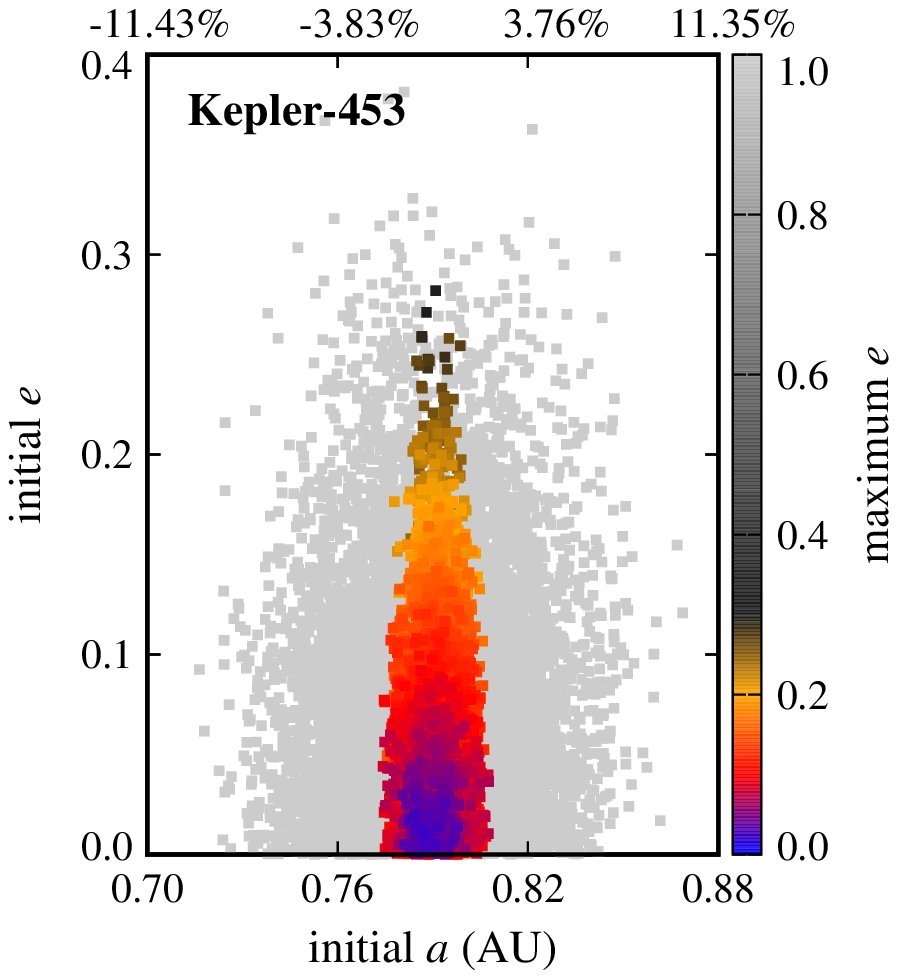}
\hskip 34pt
\includegraphics[width=0.42\textwidth]{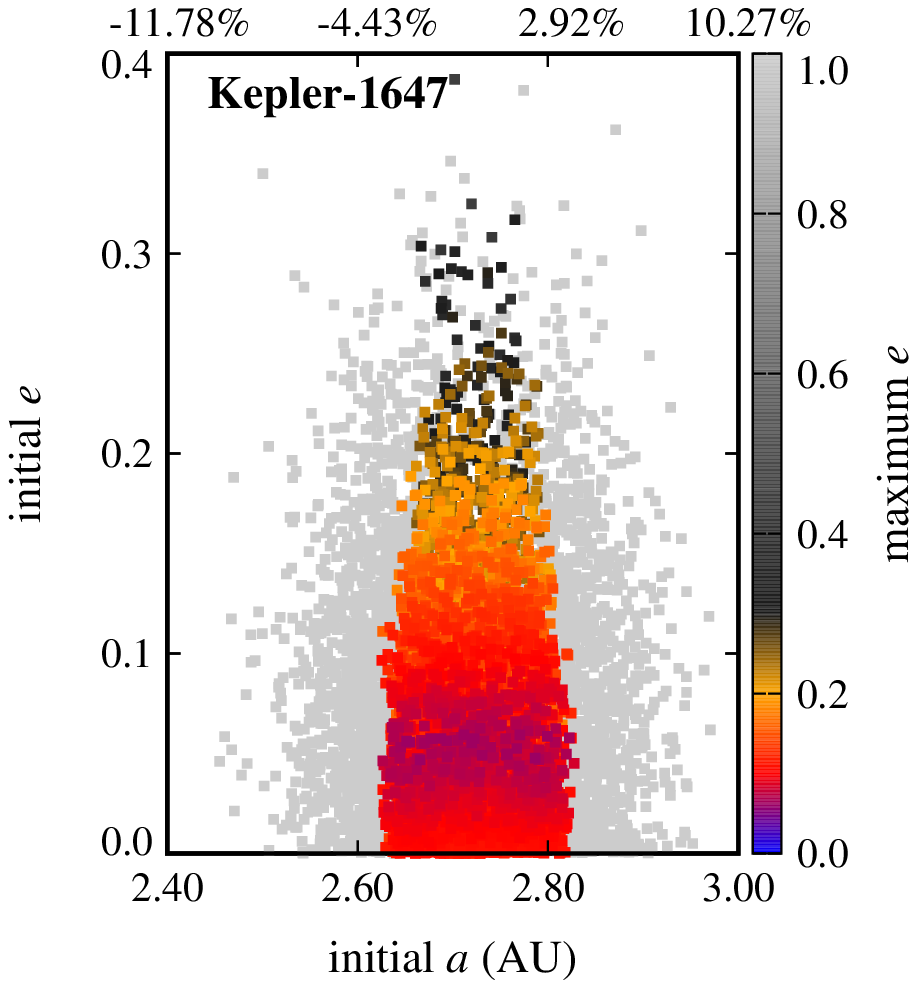}}
\caption{Maximum eccentricity $(e_{\rm max})$ of Trojan planets with respect to
their initial semimajor axes (centered on the CBP semimajor axis) and their initial eccentricities.
The top horizontal axis shows the percent deviation from the CBP semimajor axis.}
\end{figure}

\clearpage
\begin{figure}[h]
\centering
\includegraphics[width=0.42\textwidth]{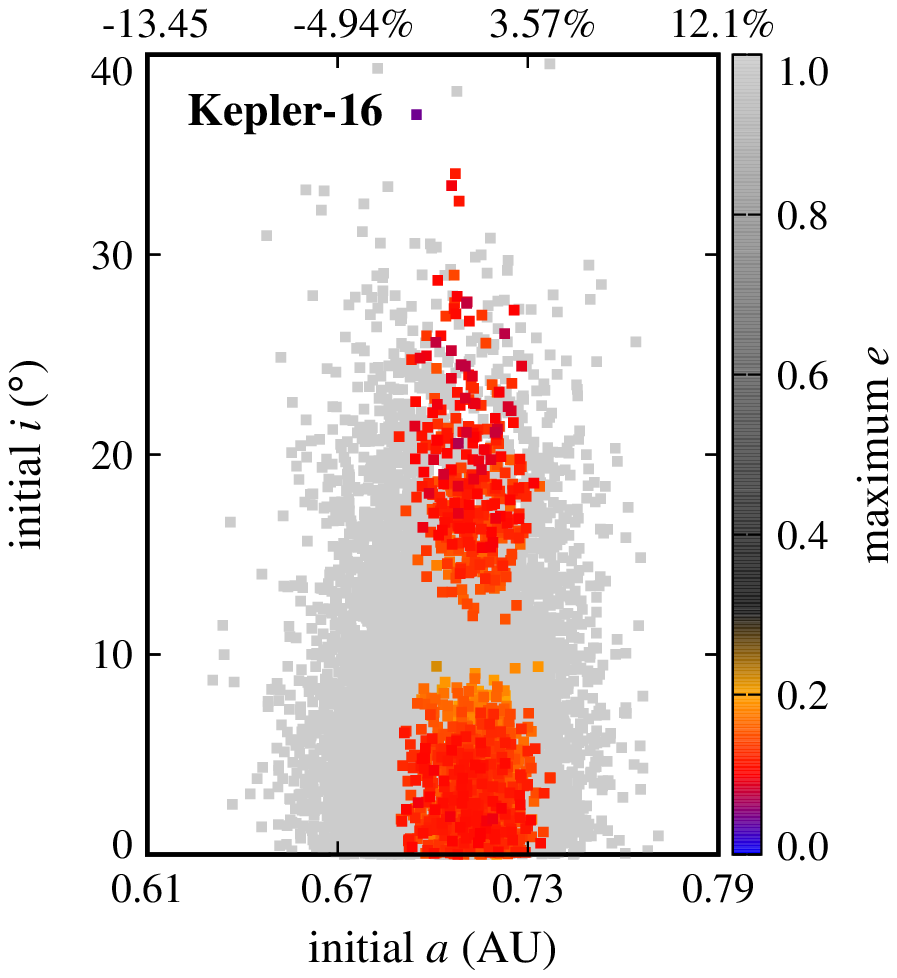}
\hskip 34pt
\includegraphics[width=0.42\textwidth]{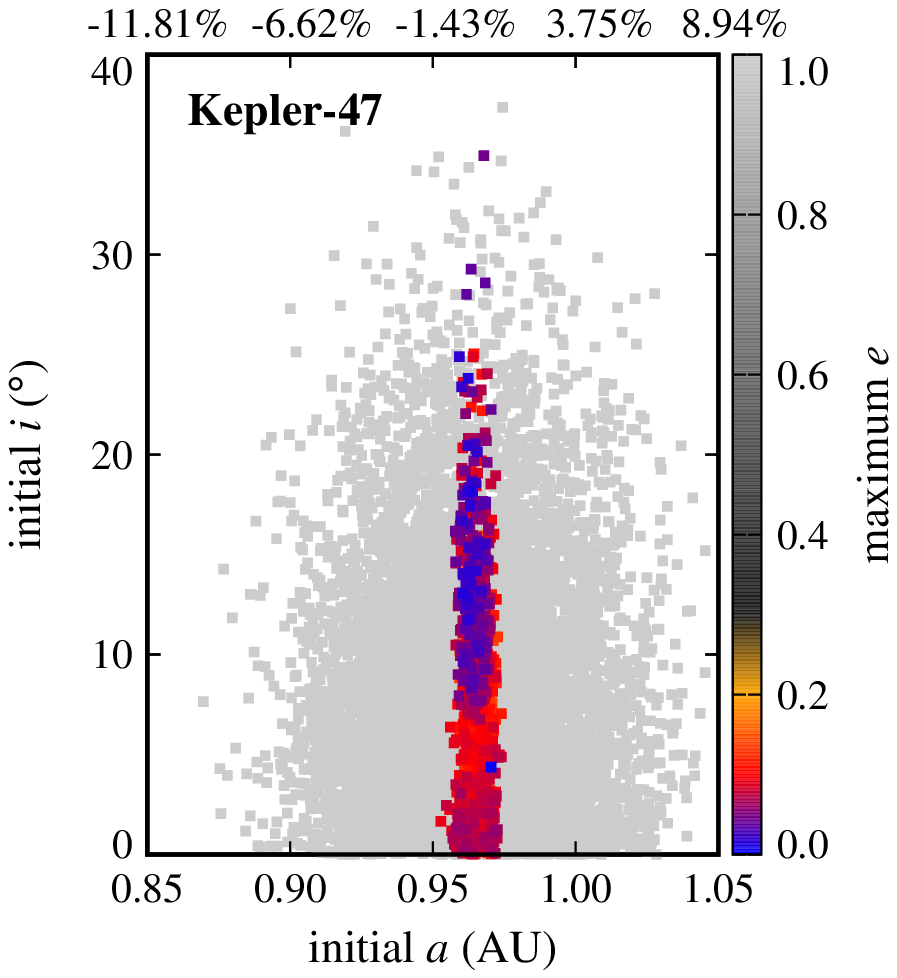}
\center{
\vskip -10pt
\includegraphics[width=0.42\textwidth]{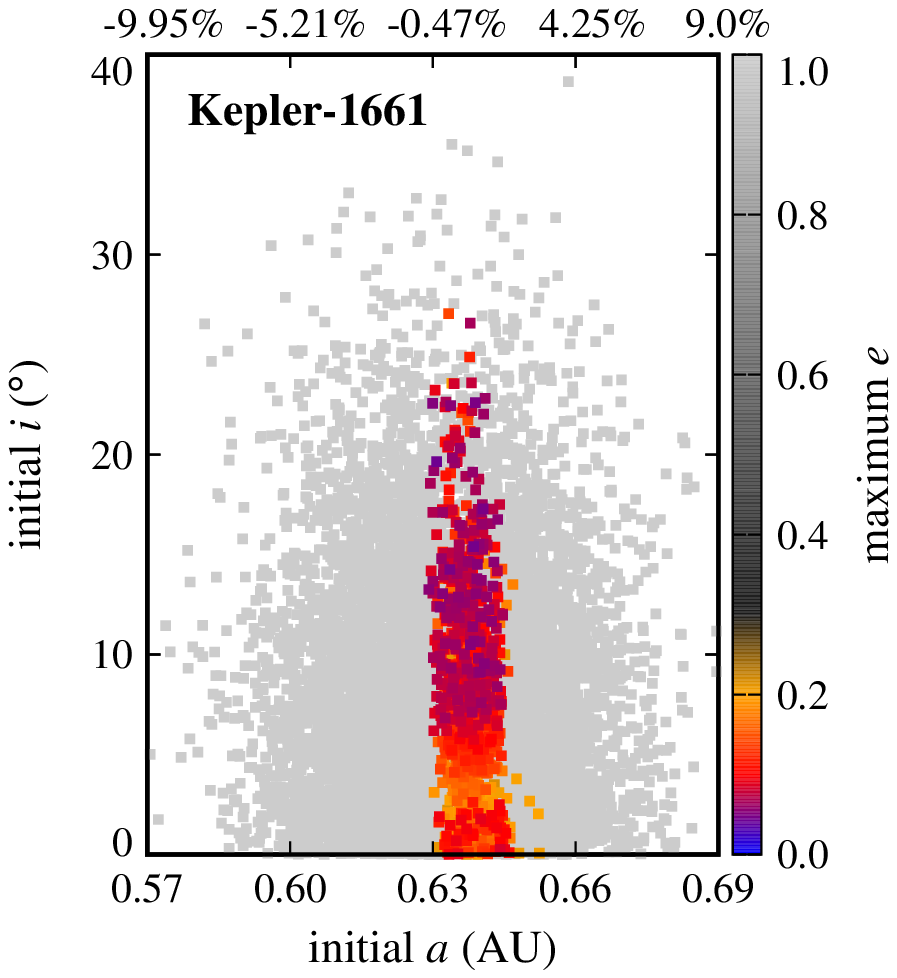}}
\center{
\vskip -10pt
\includegraphics[width=0.42\textwidth]{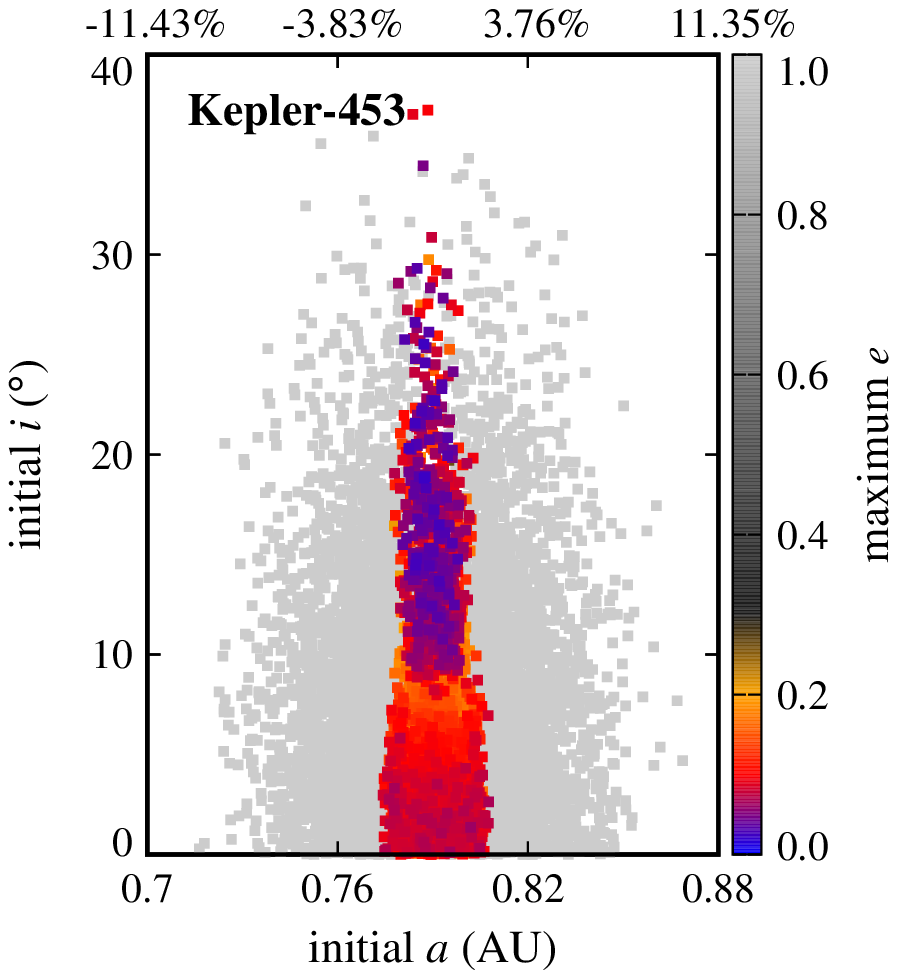}
\hskip 34pt
\includegraphics[width=0.42\textwidth]{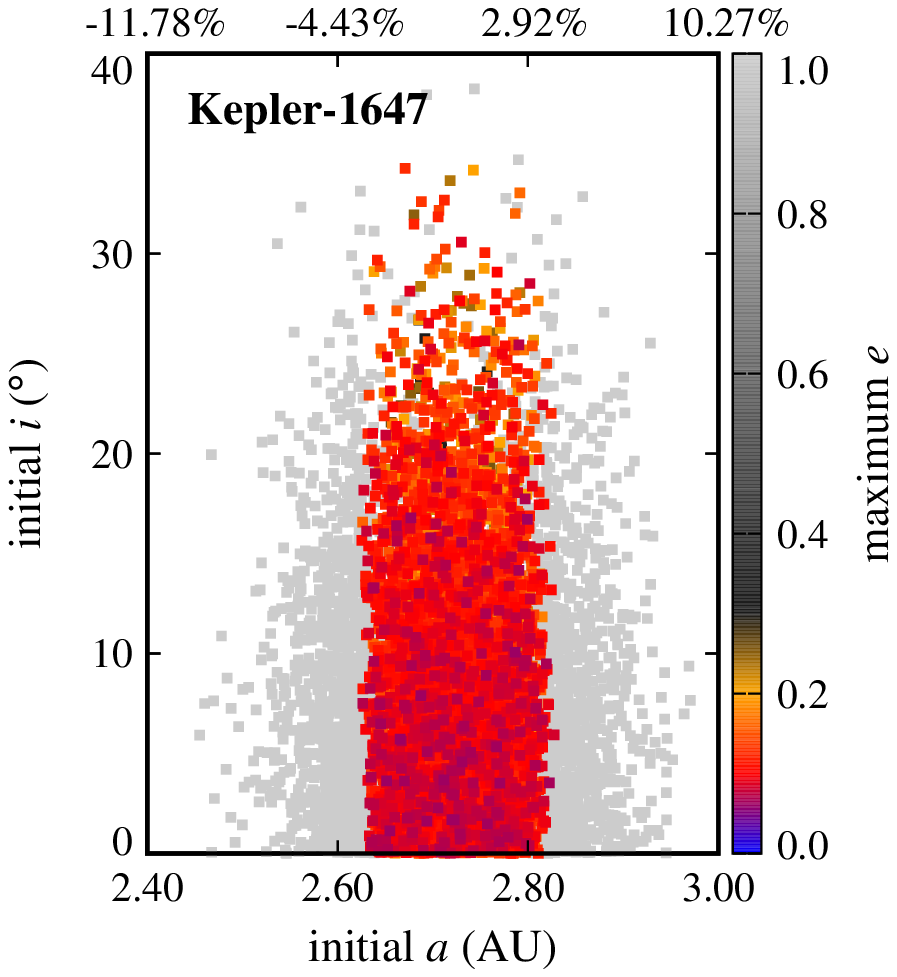}}
\caption{Maximum eccentricity $(e_{\rm max})$ of Trojan planets with respect to their initial semimajor axes 
(centered on the CBP semimajor axis) and their initial orbital inclinations.
The top horizontal axis shows the percent deviation from the CBP semimajor axis.}
\end{figure}

\clearpage
\begin{figure}[h]
\center
\includegraphics[width=0.4\textwidth]{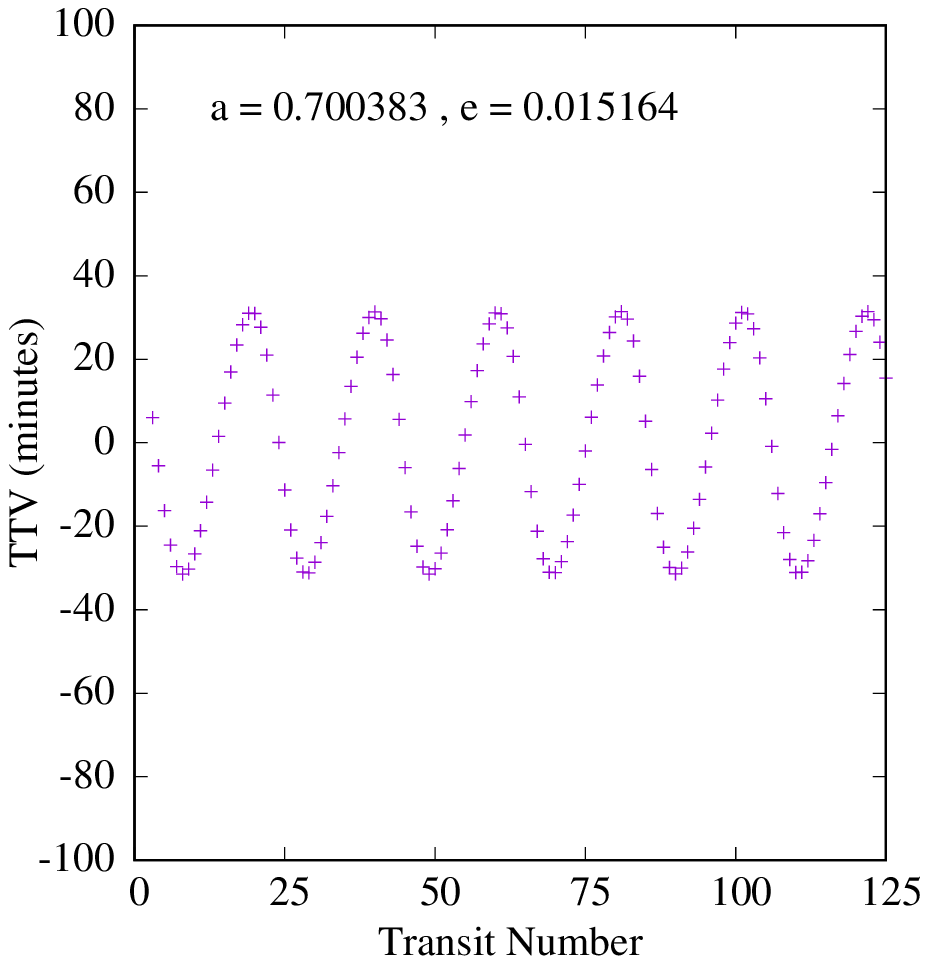}
\includegraphics[width=0.4\textwidth]{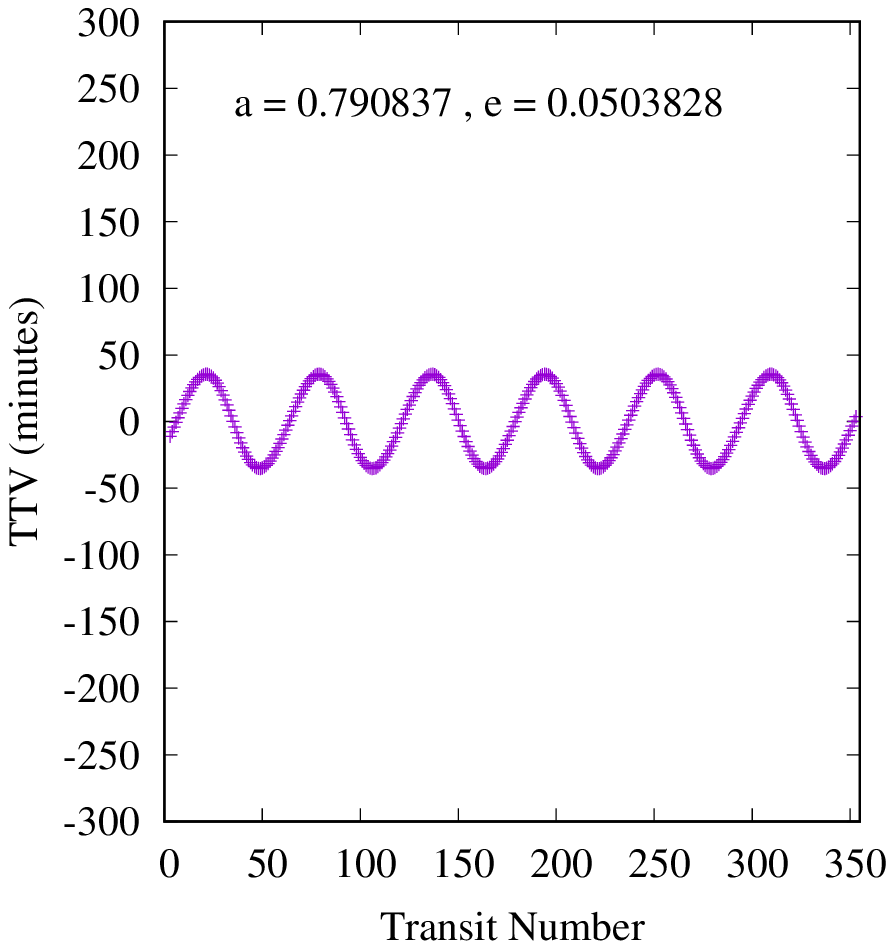}
\vskip 5pt
\includegraphics[width=0.4\textwidth]{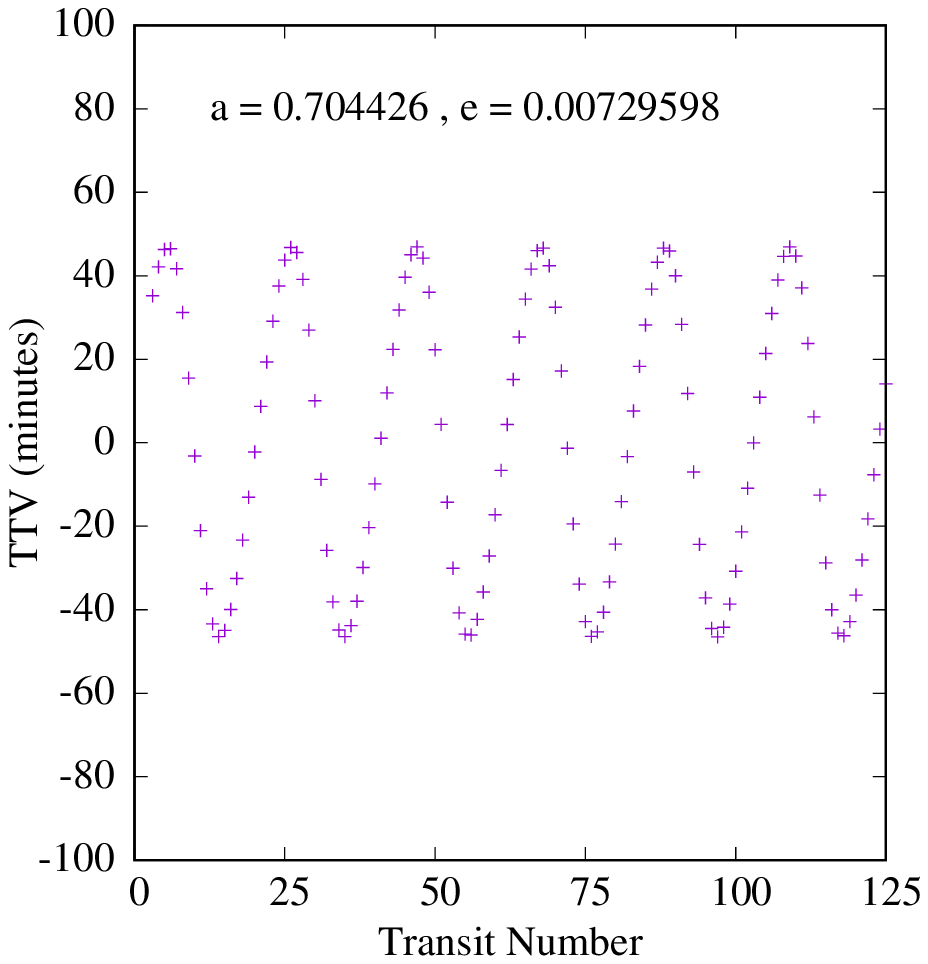}
\includegraphics[width=0.4\textwidth]{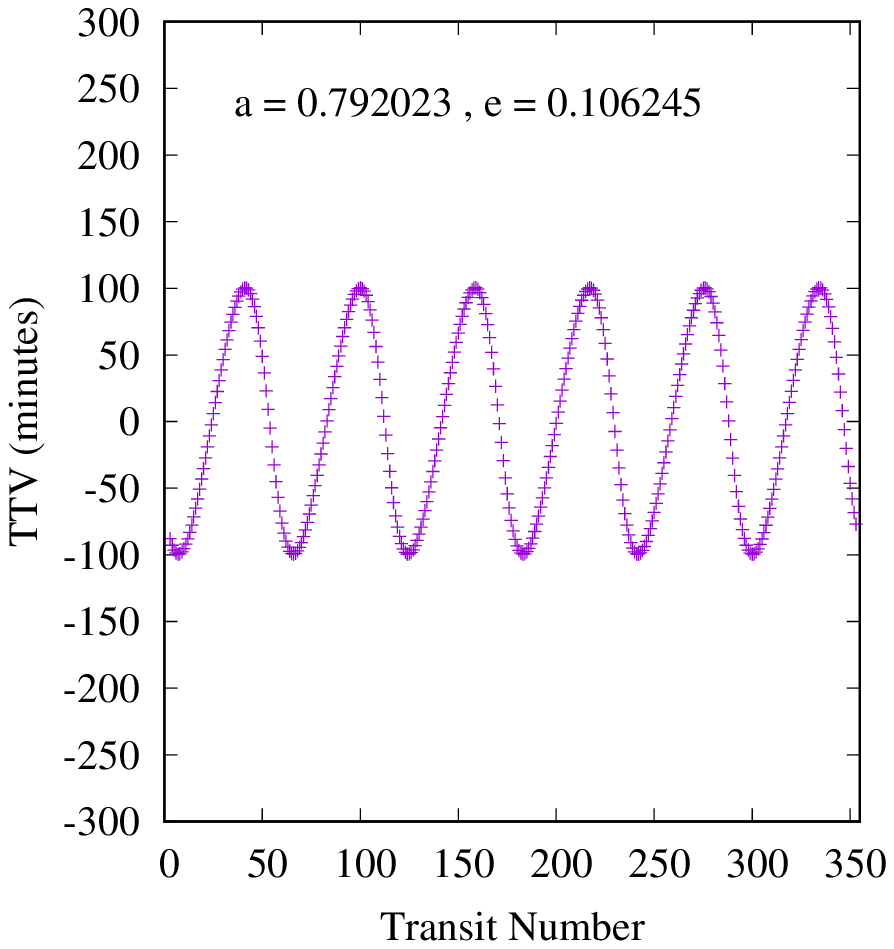}
\vskip 5pt
\includegraphics[width=0.4\textwidth]{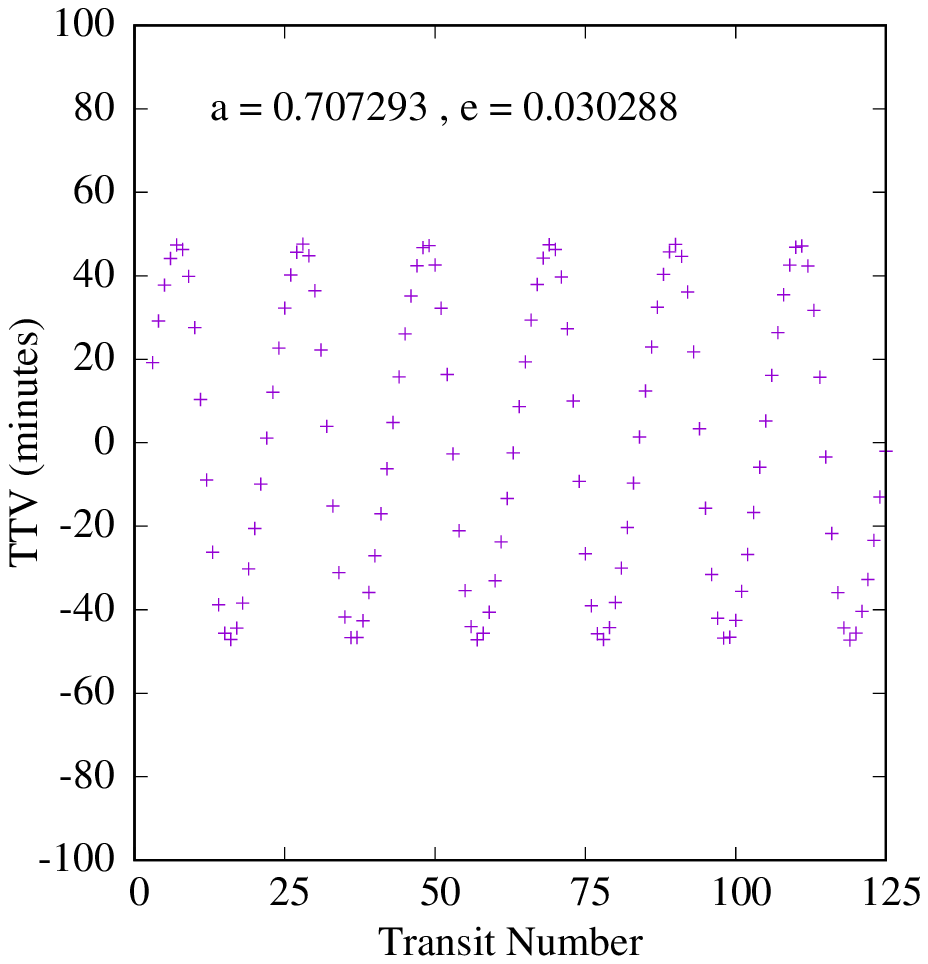}
\includegraphics[width=0.4\textwidth]{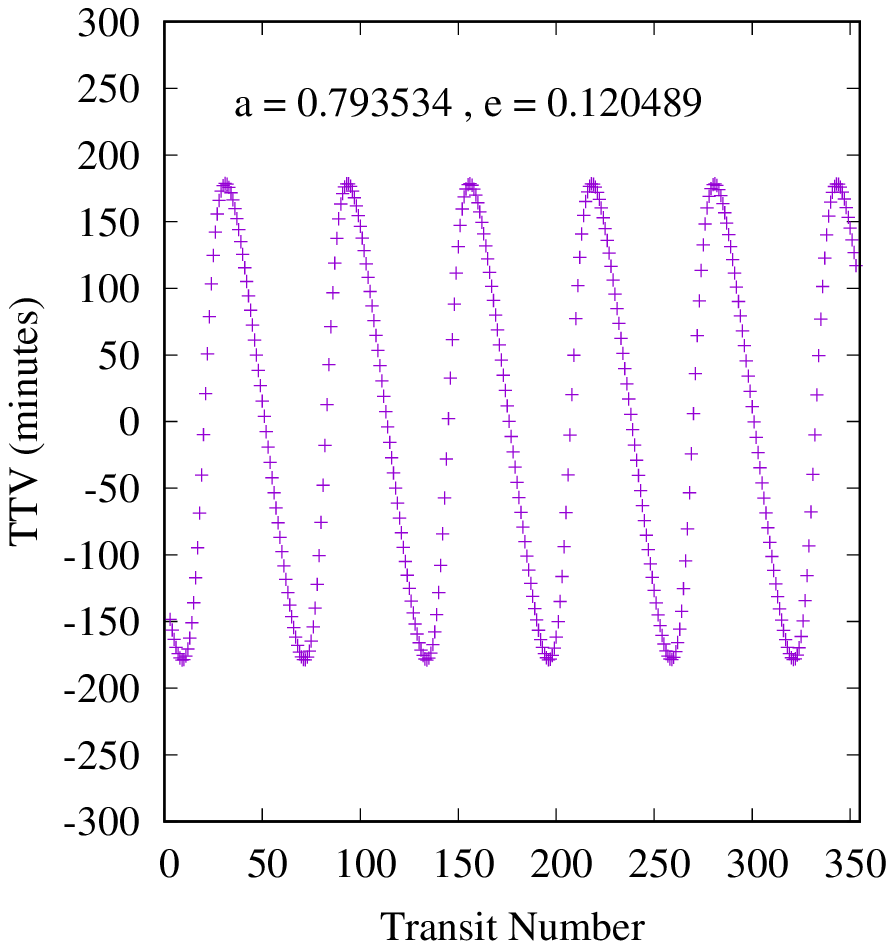}
\caption{Sample TTVs of Kepler-16b (left) and Kepler-453b (right)
due to an Earth-mass Trojan planet. The initial semimajor axis $(a)$ and eccentricity $(e)$ of 
the Trojan planet are shown in each panel.}
\end{figure}

\clearpage
\begin{figure}[h]
\center
\includegraphics[width=0.4\textwidth]{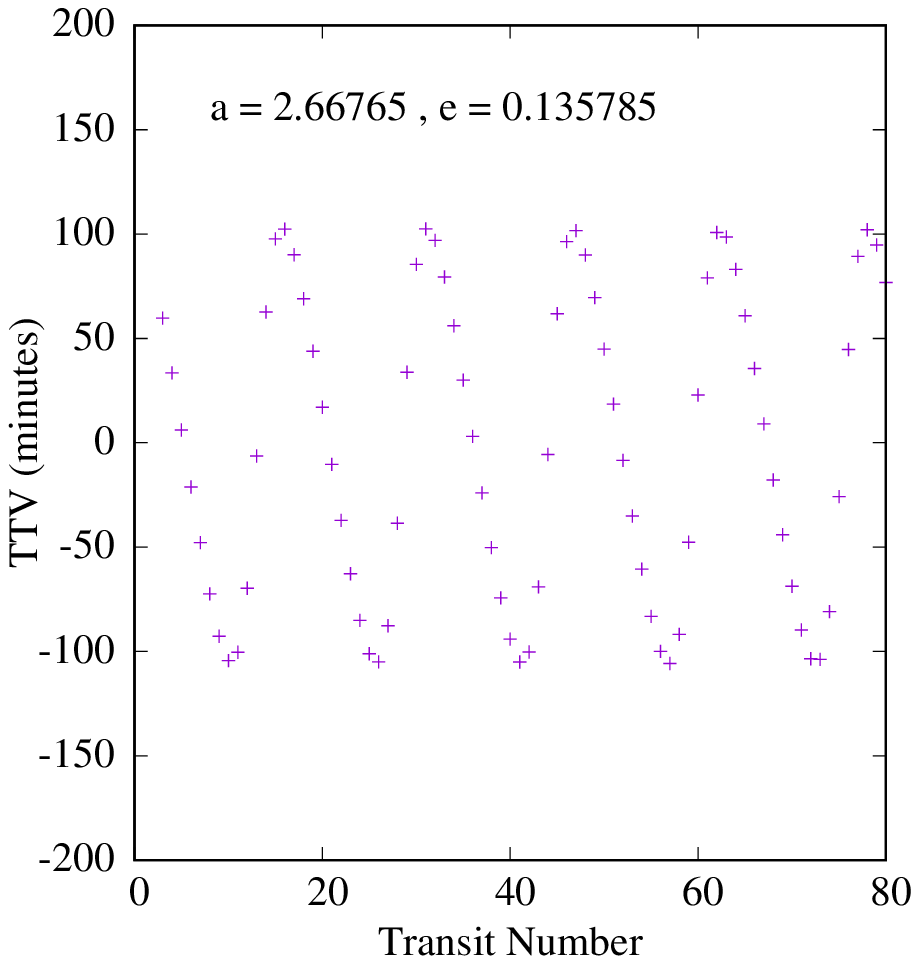}
\includegraphics[width=0.4\textwidth]{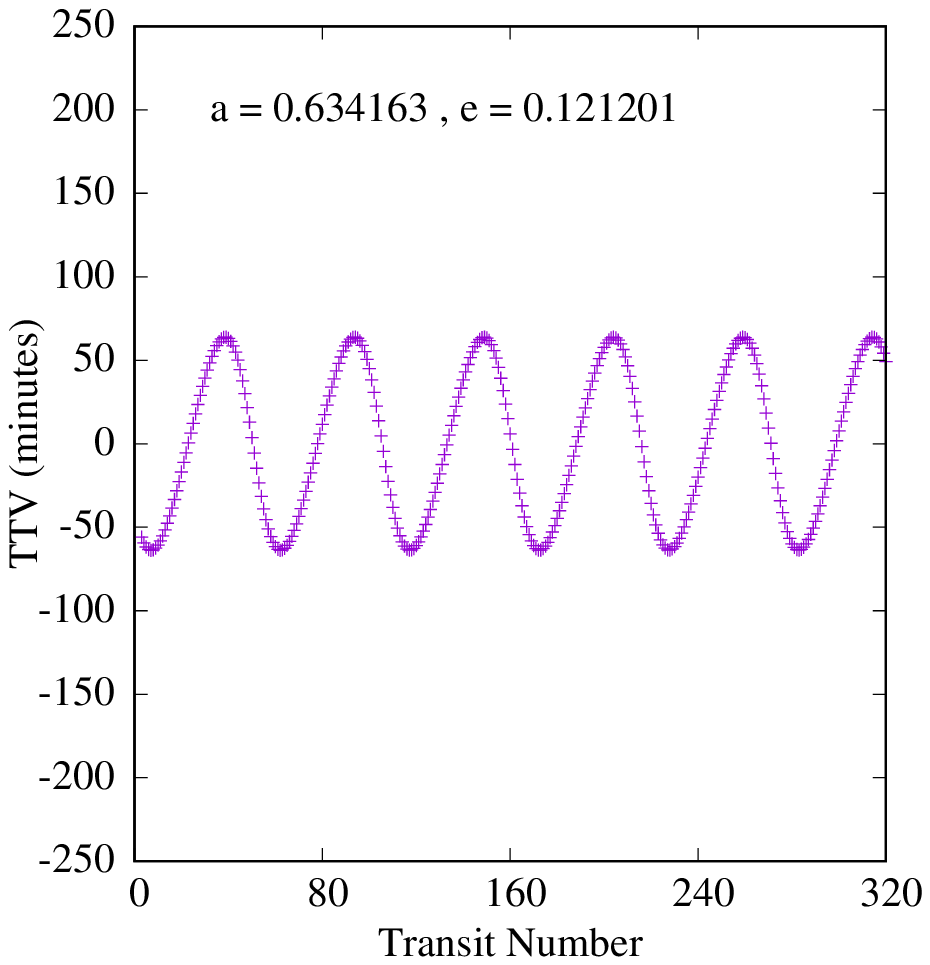}
\vskip 5pt
\includegraphics[width=0.4\textwidth]{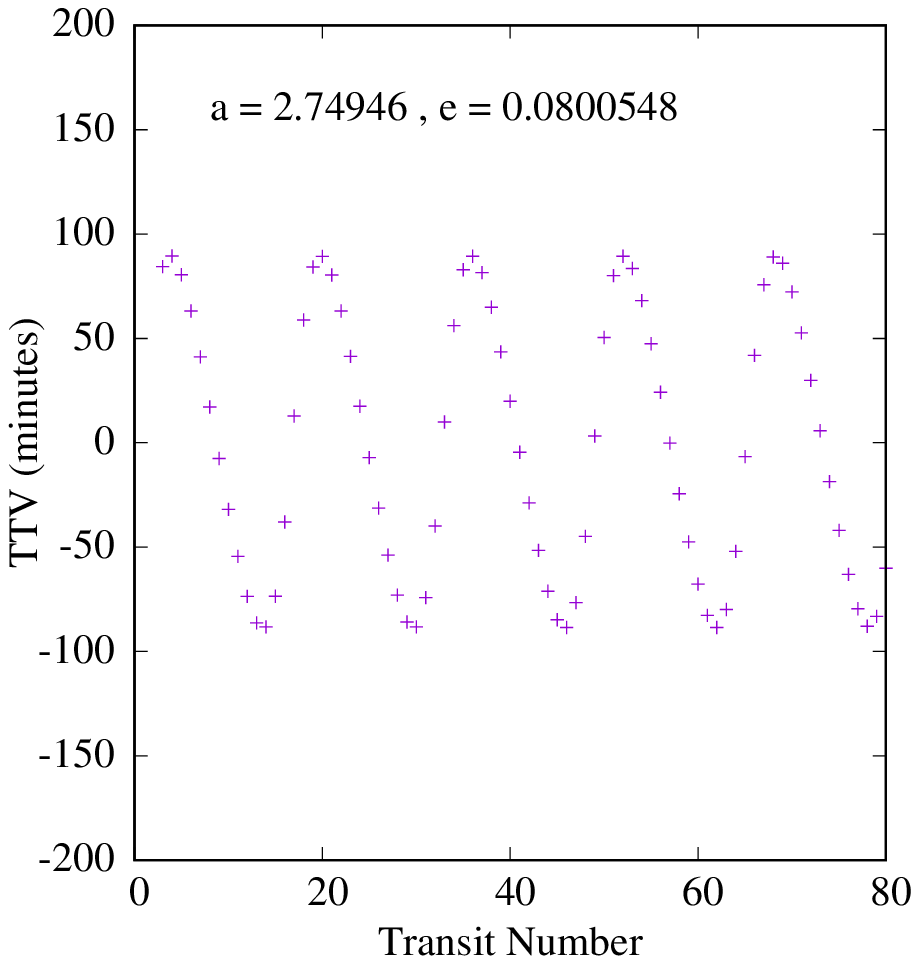}
\includegraphics[width=0.4\textwidth]{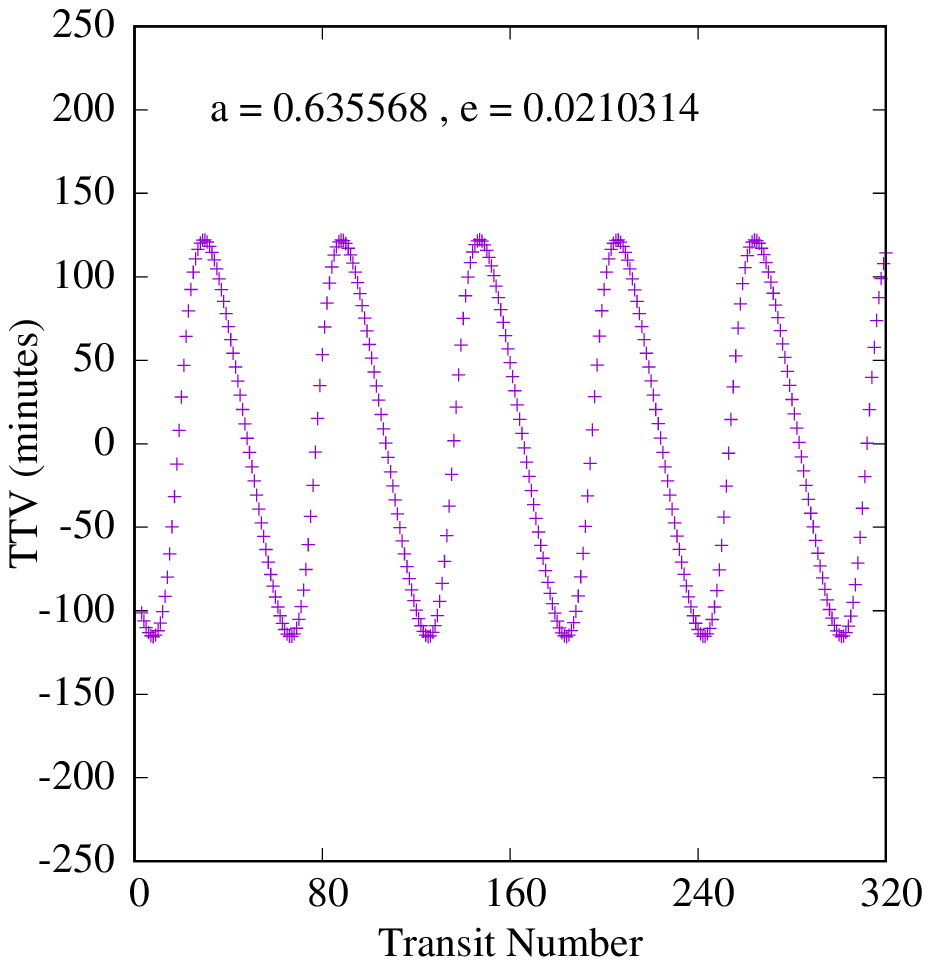}
\vskip 5pt
\includegraphics[width=0.4\textwidth]{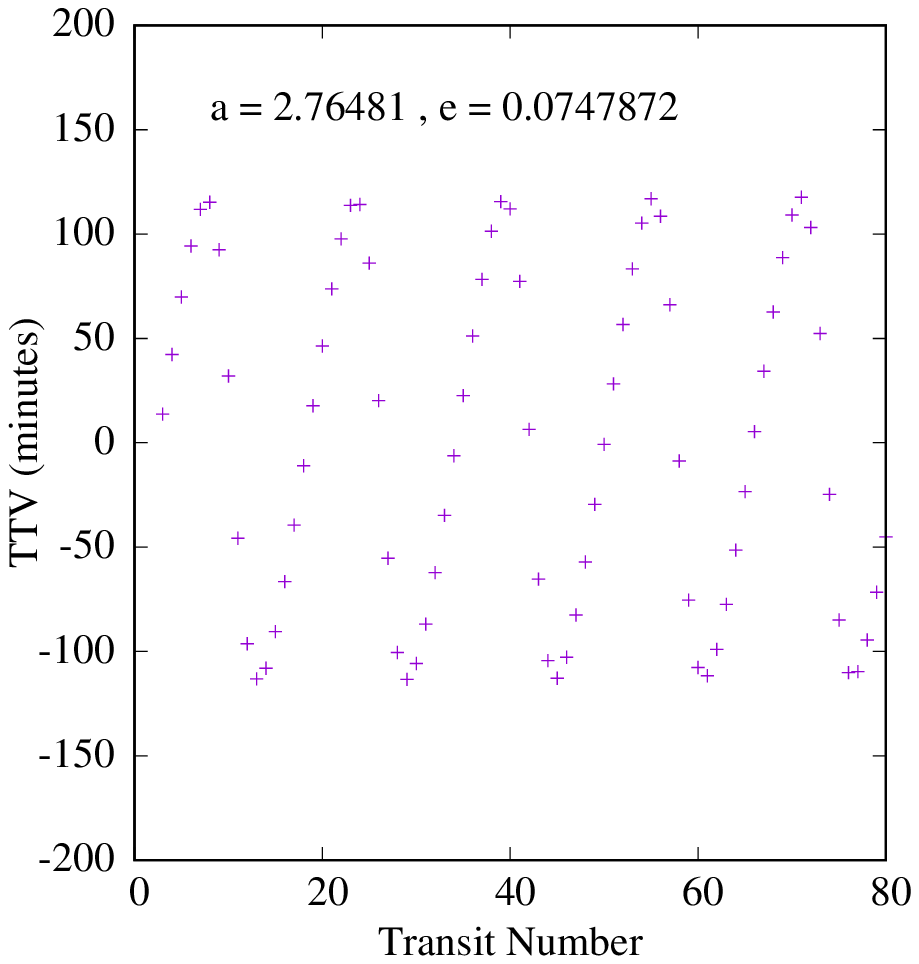}
\includegraphics[width=0.4\textwidth]{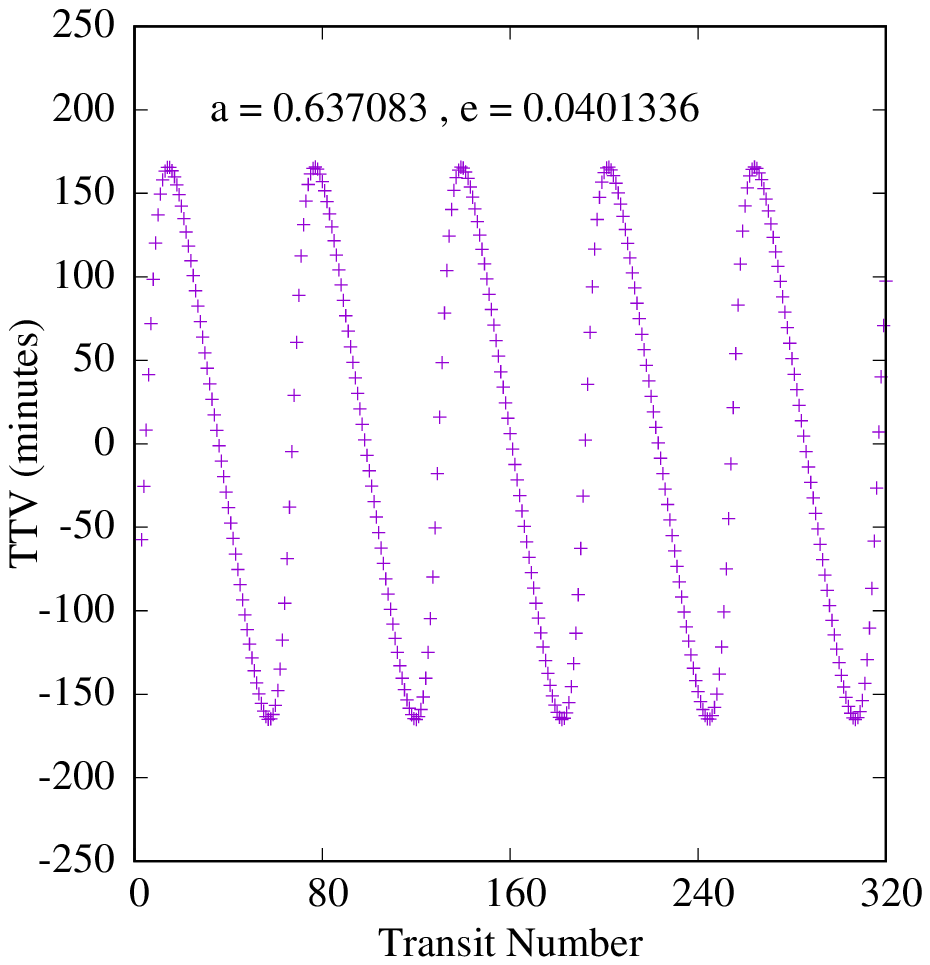}
\caption{Sample TTVs of Kepler-1647b (left) and Kepler-1661b (right)
due to an Earth-mass Trojan planet. The initial semimajor axis $(a)$ and eccentricity $(e)$ of 
the Trojan planet are shown in each panel.}
\end{figure}

\clearpage
\begin{figure}[h]
\center
\includegraphics[width=0.4\textwidth]{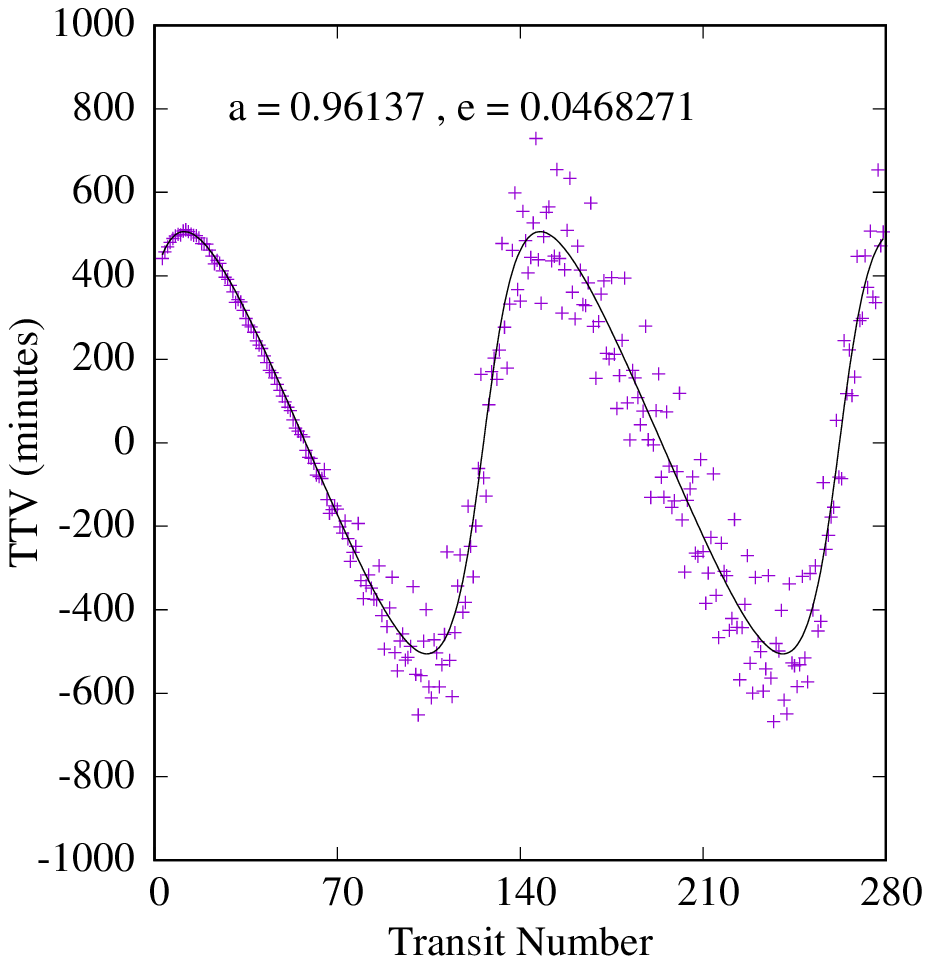}
\includegraphics[width=0.4\textwidth]{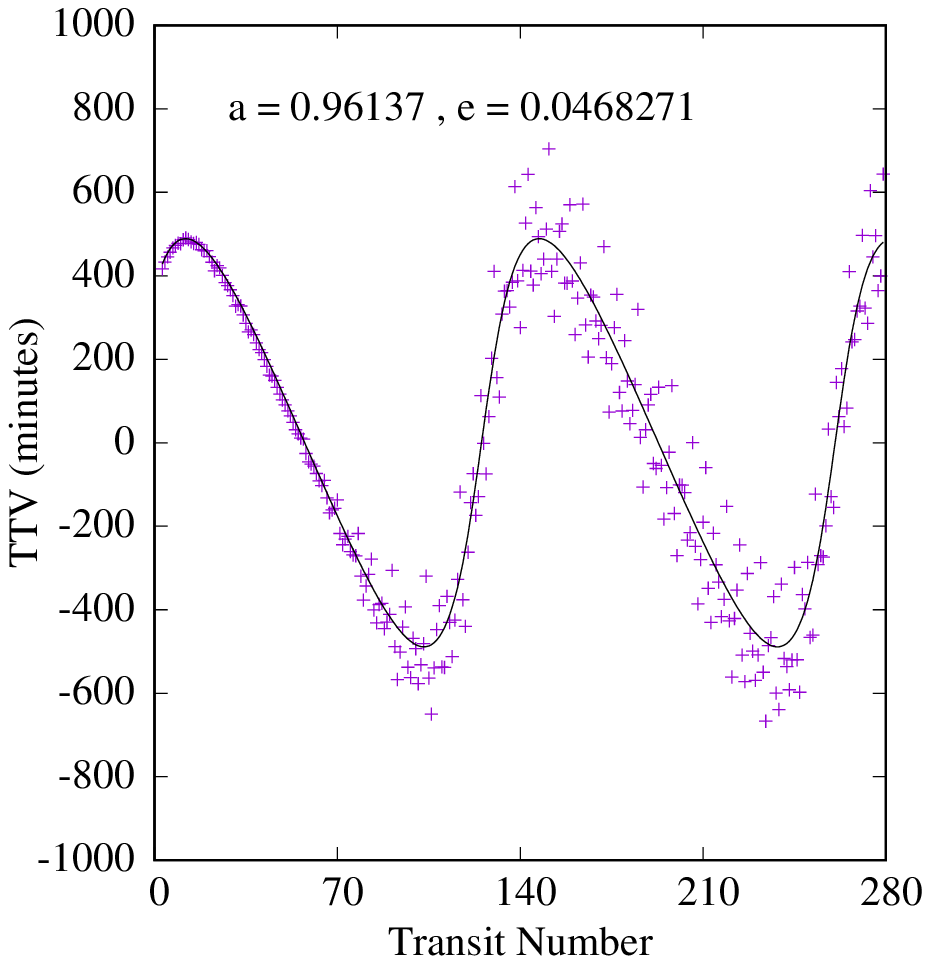}
\vskip 5pt
\includegraphics[width=0.4\textwidth]{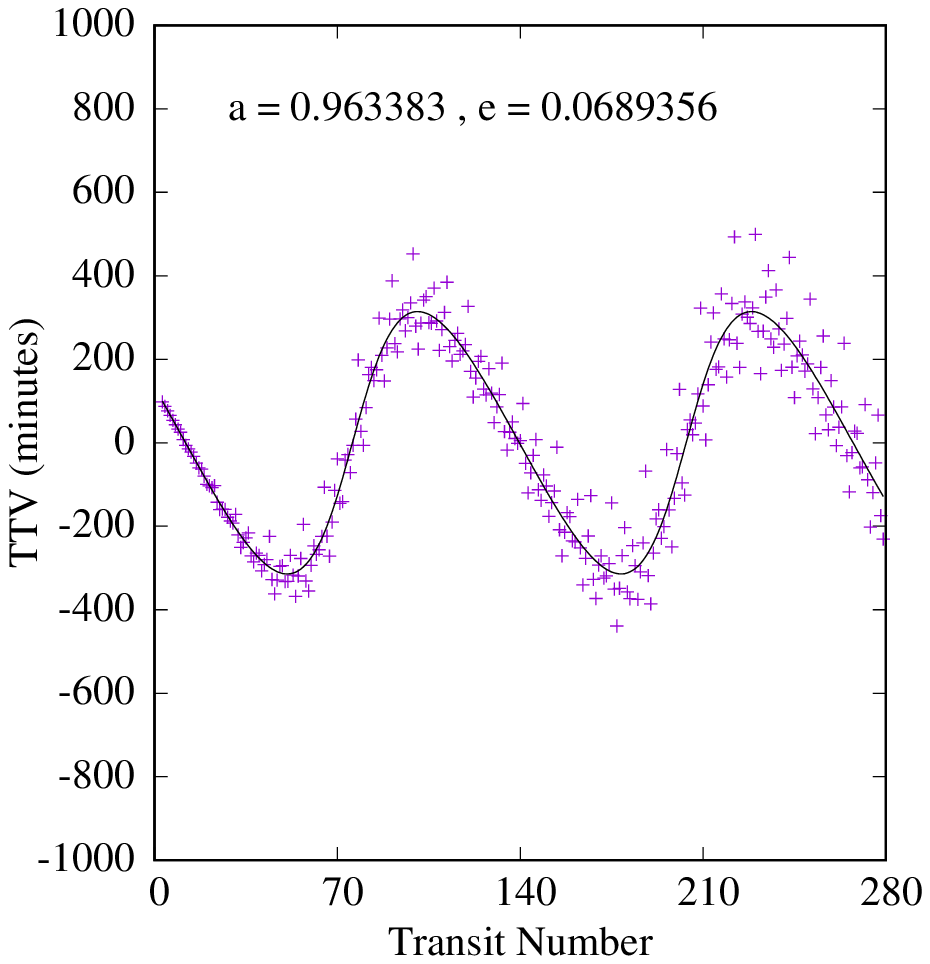}
\includegraphics[width=0.4\textwidth]{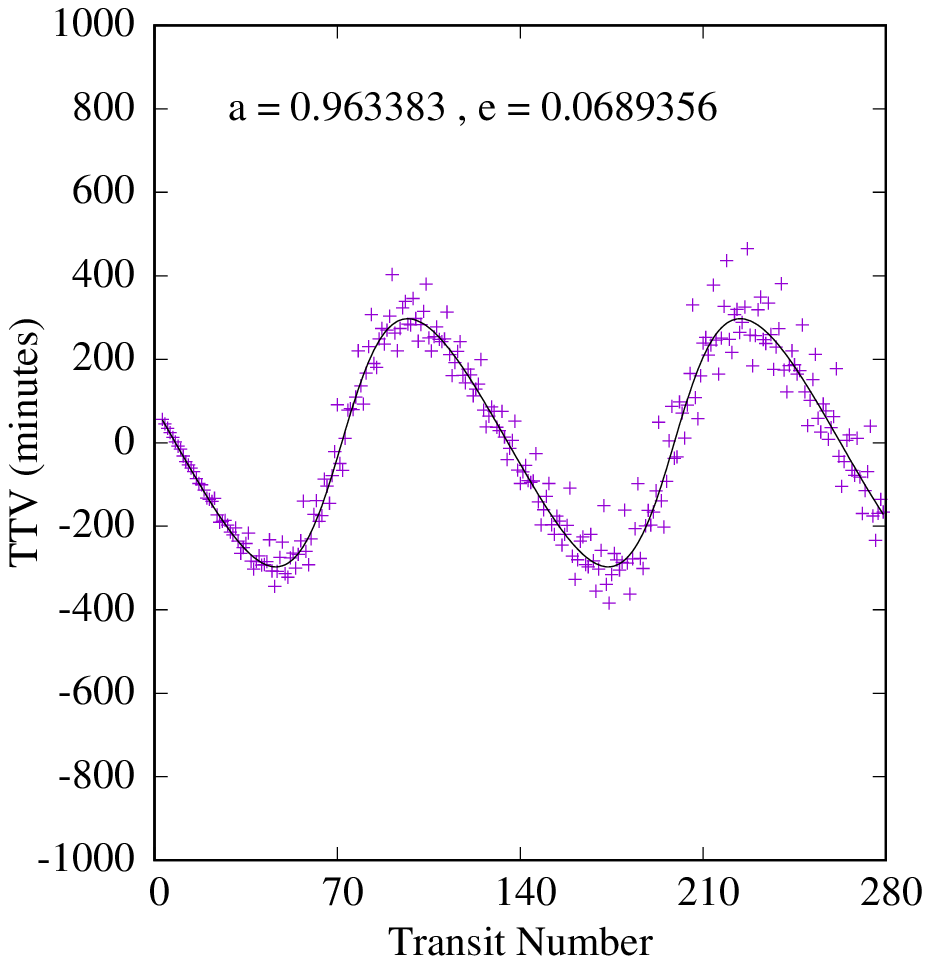}
\vskip 5pt
\includegraphics[width=0.4\textwidth]{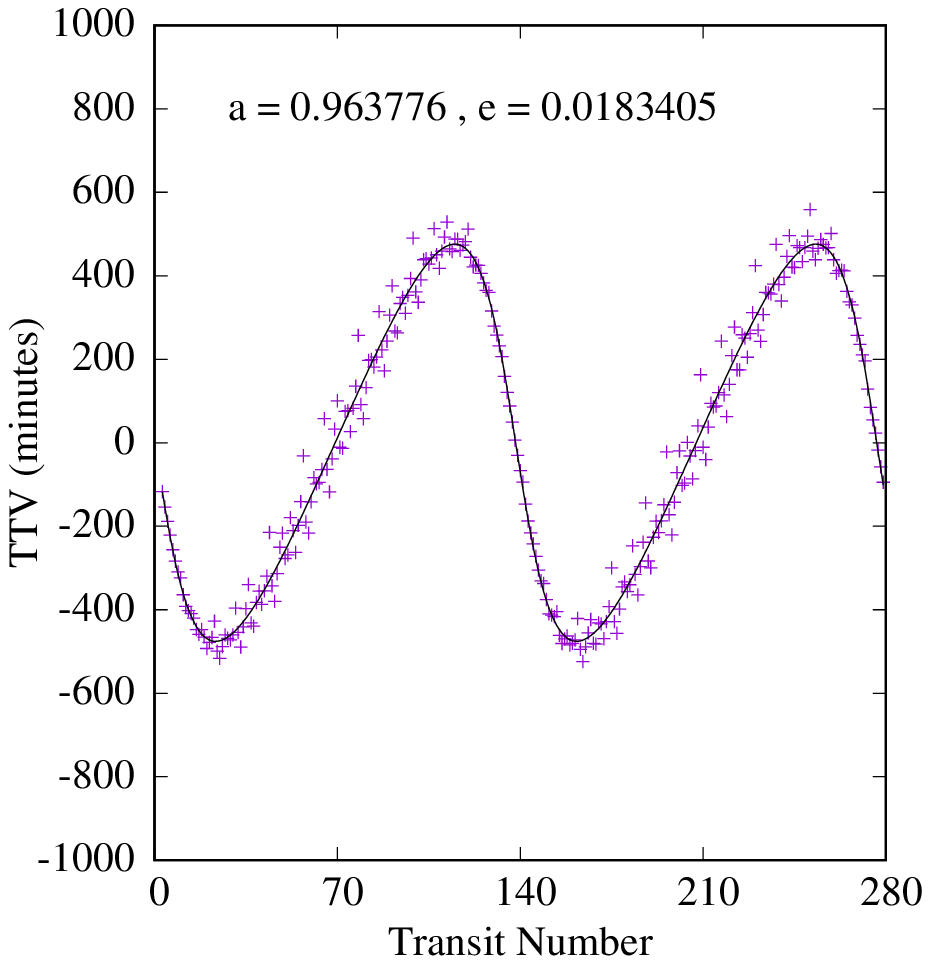}
\includegraphics[width=0.4\textwidth]{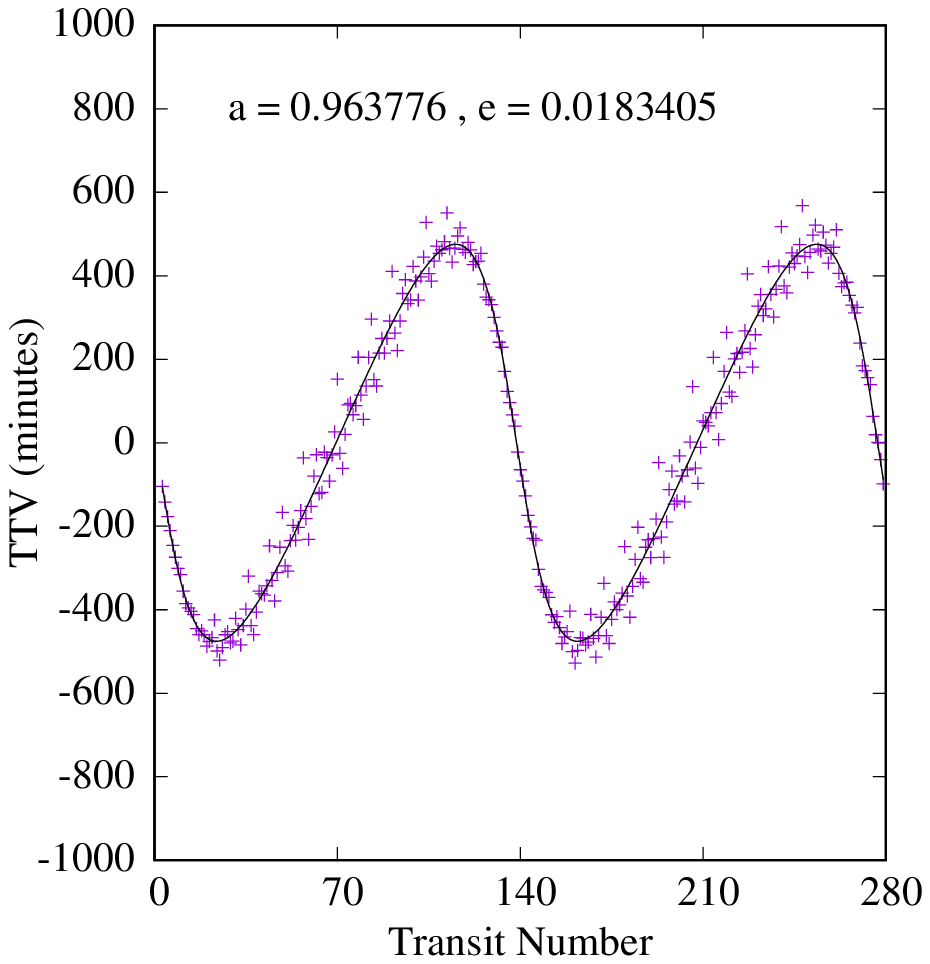}
\vskip -10pt
\caption{Sample TTVs of the CBP Kepler-47c for the low mass estimate of the inner planet Kepler-47b (left) 
and its highest mass (right) due to an Earth-mass Trojan planet. The initial semimajor axis $(a)$ 
and eccentricity $(e)$ of the Trojan planet are shown in each panel. The solid curve represents a titled-sine fit 
to the TTV data. The scatter around the fitted curve is caused by Kepler-47b and as shown by the figure,
is almost independent of the choice of its mass.}
\end{figure}

\clearpage
\begin{deluxetable}{lcccccccc}
\tablenum{1}
\tablecaption{Mass and orbital elements of {\it Kepler} HZ circumbinary systems.
\label{table1}}
\tablecolumns{9}
\tablewidth{0pt}
\tablehead{
{Binary} & \colhead{Primary} & \colhead{Secondary} & \colhead{$a$} & \colhead{$e$} & \colhead{$i$} & \colhead {$\omega$} & 
\colhead{$\Omega$} & \colhead{M.A.} \\
\colhead{} & \colhead{ $(M_\odot)$}  & \colhead{ $(M_\odot)$} & \colhead{(AU)} & \colhead{} & \colhead{(deg.)} & \colhead{(deg.)} 
& \colhead{(deg.)} & \colhead{(deg.)}}
\startdata
Kepler-16    & 0.6897  & 0.20255  & 0.22431    & 0.15944  & 0   & 263.464   & 0   & 188.888   \\
Kepler-47     & 0.957   & 0.342    & 0.08145    & 0.0288   & 0   & 226.3     & 0   & 310.8     \\
Kepler-453    & 0.944   & 0.1951   & 0.18539    & 0.0524   & 0   & 263.05    & 0   & 182.0931  \\
Kepler-1647   & 1.2207  & 0.9678   & 0.1276     & 0.1602   & 0   & 300.5442  & 0   & 90.62     \\
Kepler-1661   & 0.841   & 0.262    & 0.187      & 0.112    & 0   & 36.4      & 0   & 253.983   \\
\enddata
\end{deluxetable}

\clearpage
\begin{deluxetable}{lccccccc}
\tablenum{2}
\tablecaption{Mass and orbital elements of {\it Kepler} HZ circumbinary planets.
\label{table2}}
\tablecolumns{8}
\tablewidth{0pt}
\tablehead{
{Planet} & \colhead{Mass} & \colhead{$a$} & \colhead{$e$} & \colhead{$i$} & \colhead {$\omega$} & 
\colhead{$\Omega$} & \colhead{M.A.} \\
\colhead{} & \colhead{$(M_{\oplus})$} & \colhead{(AU)} & \colhead{} & \colhead{(deg.)} & \colhead{(deg.)} &
\colhead{(deg.)} & \colhead{(deg.)}}
\startdata
Kepler-16b     & 106      & 0.7048     & 0.0069   & 0      & 318       & 0       & 148.507   \\
Kepler-47b     & 2.073    & 0.2877     & 0.021    & 0.166  & 48.6      & 0       & 13.32 \\
Kepler-47c     & 3.17    & 0.9638     & 0.044    & 1.38   & 306       & 0       & 165.4     \\
Kepler-47d     & 19.03   & 0.6992     & 0.024    & 1.165  & 352       & 0       & 300.7       \\
Kepler-453b    & 16       & 0.7903     & 0.0359   & 2.258  & 185.1     & 2.103   & 113.3028  \\
Kepler-1647b   & 483.33   & 2.7205     & 0.0581   & 2.9855 & 155.0464  & -2.0393 & 301.37     \\
Kepler-1661b   & 17       & 0.633      & 0.057    & 0.93   & 67.1         & 0.61    & 37.441   \\
\enddata
\end{deluxetable}

\clearpage
\begin{deluxetable}{lcccccc}
%\tabletypesize{\scriptsize}
\tablecaption{Initial and final population of Trojan planets in the $L_4$ (leading) and $L_5$ (trailing)
Lagrangian points
\label{table3}}
\tablenum{3}
\tablewidth{0pt}
%\tablecolumns{2}
\tablehead{\colhead{} & \colhead{} & \multicolumn{2}{c}{Initial Population} & & \multicolumn{2}{c}{Stable Population} \\
%\colhead{$\alpha$ Cen AB} & \multicolumn{2}{c}{Narrow HZ} & \multicolumn{2}{c}{Nominal HZ} \\
\cline{3-4}
\cline{6-7} 
\colhead {} & \colhead {} & \colhead{$L_4$} & \colhead {$L_5$} &   \colhead{} & \colhead{$L_4$} & \colhead {$L_5$} 
%\colhead {} & \colhead{inner} & \colhead {outer} &  \colhead{inner} & \colhead {outer}
}
\startdata
Kepler-16b    & & 4937 & 5063  & & 356  & 765  \\
Kepler-47c    & & 4987 & 5013  & & 484  & 412  \\
Kepler-453b   & & 5022 & 4978  & & 1660 & 1719 \\
Kepler-1647b  & & 4944 & 4868  & & 3002 & 3285 \\
Kepler-1661b  & & 4975 & 5025  & & 520  & 501  \\
\enddata
\end{deluxetable}

\clearpage
\begin{deluxetable}{lcccc}
\tablenum{4}
\tablecaption{Dynamical state of Trojan planets at the end of the integrations. Approximately
10,000 planets were integrated in each system. 
\label{table4}}
\tablecolumns{5}
\tablewidth{0pt}
\tablehead{
{Planet} & \colhead{Collided with} & \colhead{Ejected} & \colhead{Collided with} & \colhead{Stable}\\
\colhead{} & \colhead{Planet} & \colhead{} & \colhead{Star} & \colhead{}}
\startdata
Kepler-16b     & 5392   & 3291   & 16   & 1301  \\
Kepler-47b     & 29     & 0      & 0    &  -     \\
Kepler-47c     & 8921   & 0      & 0    & 896   \\
Kepler-47d     & 154    & 0      & 0    &  -     \\
Kepler-453b    & 6618   & 3      & 0    & 3379  \\
Kepler-1647b   & 3523   & 2      & 0    & 6287  \\
Kepler-1661b   & 8319   & 657    & 3    & 1021  \\
\enddata
\end{deluxetable}

\clearpage
\begin{deluxetable}{lcccc}
\tablenum{5}
\tablecaption{Characteristics of the transit timing variations of {\it Kepler} HZ CBPs arranged
in order of decreasing mass.
\label{table5}}
\tablecolumns{5}
\tablewidth{0pt}
\tablehead{
{Planet} & \colhead{TTV Period} & \colhead{Maximum} & \colhead{Minimum} & \colhead{Median}\\
\colhead{} & \colhead{(Transit \#)} & \colhead{(min)} & \colhead{(min)} & \colhead{(min)}}
\startdata
Kepler-1647b            & 16   & 194   & 10   & 97   \\
Kepler-16b              & 20   & 151   & 4    & 71   \\
Kepler-1661b            & 64   & 198   & 7    & 128  \\
Kepler-453b             & 60   & 262   & 9    & 153  \\
Kepler-47c (High-mass)  & 128  & 598   & 60   & 377  \\
Kepler-47c (Low-mass)   & 128  & 598   & 60   & 390  \\
\enddata
\end{deluxetable}

\end{document}